\pdfoutput=1

\documentclass[11pt,twoside,a4paper,cmspaper,final,collab]{cms-tdr}
\def\svnVersion{433995P}\def\cmsCernNoTag{CERN-EP-2016-254}\def\cmsCernDate{\today}\def\cmsMessage{Published in the European Physical Journal C as \href{http://dx.doi.org/10.1140/epjc/s10052-017-5140-y}{\doi{10.1140/epjc/s10052-017-5140-y.}}}

\begin{document}\cmsNoteHeader{SMP-14-010}

\hyphenation{had-ron-i-za-tion}
\hyphenation{cal-or-i-me-ter}
\hyphenation{de-vices}
\RCS$Revision: 419178 $
\RCS$HeadURL: svn+ssh://svn.cern.ch/reps/tdr2/papers/SMP-14-010/trunk/SMP-14-010.tex $
\RCS$Id: SMP-14-010.tex 419178 2017-08-02 08:23:21Z fabiocos $
\newlength\ghmFigWidth
\setlength\ghmFigWidth{0.48\textwidth}
\newlength\ghmFigWidthTwo
\ifthenelse{\boolean{cms@external}}{\setlength\ghmFigWidthTwo{0.42\textwidth}}{\setlength\ghmFigWidthTwo{0.48\textwidth}}
\newlength\cmsFigWidth
\ifthenelse{\boolean{cms@external}}{\setlength\cmsFigWidth{0.85\columnwidth}}{\setlength\cmsFigWidth{0.4\textwidth}}
\ifthenelse{\boolean{cms@external}}{\providecommand{\cmsLeft}{top}}{\providecommand{\cmsLeft}{left}}
\ifthenelse{\boolean{cms@external}}{\providecommand{\cmsRight}{bottom}}{\providecommand{\cmsRight}{right}}
\newcommand{\POWHEGBOX} {{\textsc{powheg-box}}\xspace}
\newcommand{\AMCATNLO} {\textsc{MadGraph5}\_{\allowbreak}a\textsc{MC@NLO}\xspace}
\newcommand{\PYTHIAS} {{\textsc{pythia}6}\xspace}
\newcommand{\PYTHIAE} {{\textsc{pythia}8}\xspace}
\renewcommand{\kt}{\ensuremath{k_\mathrm{t}}\xspace}
\newcommand{\Zb}{\ensuremath{$\PZ$+\PQb}\xspace}
\newcommand{\Zob}{\ensuremath{\PZ(1\PQb)}\xspace}
\newcommand{\Ztb}{\ensuremath{\PZ(2\PQb)}\xspace}
\cmsNoteHeader{SMP-14-010}
\title{Measurements of the associated production of a Z boson and b jets in pp collisions at \ifthenelse{\boolean{cms@external}}{\boldmath{$\sqrt{s}$} = 8\TeV}{$\sqrt{s} = 8\TeV$}}

\date{\today}
\titlerunning{Measurements of the associated production of a Z boson and b jets in pp collisions at $\sqrt{s} = 8\TeV$}
\abstract{
Measurements of the associated production of a \PZ boson with at least
one jet originating from a b quark in proton-proton collisions
at $\sqrt{s} = 8\TeV$ are presented. Differential cross sections are
measured with data collected by the CMS experiment corresponding to an
integrated luminosity of 19.8\fbinv. \PZ bosons are reconstructed
through their decays to electrons and muons. Cross sections are
measured as a function of observables characterizing the kinematics of
the \PQb jet and the \PZ boson. Ratios of differential cross
sections for the associated production with at least one \PQb jet
to the associated production with any jet are also presented.  The
production of a \PZ boson with at least two \PQb jets is investigated,
and differential cross sections are measured for the dijet system.
Results are compared to theoretical predictions, testing two different
flavour schemes for the choice of initial-state partons.
}
\hypersetup{%
pdfauthor={CMS Collaboration},%
pdftitle={Measurements of the associated production of a Z boson and b jets in pp collisions at sqrt(s) = 8 TeV},%
pdfsubject={CMS},%
pdfkeywords={CMS, physics, standard model}}
\maketitle
\newpage{}
\section{Introduction}
{\tolerance=1200
The associated production of vector bosons and jets (V+jets) in
hadronic collisions is a large background source in measurements of
several standard model (SM) processes, Higgs boson studies, and many
searches for physics beyond the SM.  Its description constitutes an
important benchmark for perturbative quantum chromodynamics (pQCD)
predictions. Differential cross sections as a function of kinematic
observables characterizing V+jets topologies are sensitive to the
contributions from both the hard scattering process and the associated
soft QCD radiation, as well as to the parton distribution functions
(PDFs). Among the V+jets processes, the case in which a
$\PZ/\gamma^{\ast}$ boson is produced in association with $\PQb$
quarks, $\Pp\Pp\to \PZ +({\geq}1\PQb)$, hereafter denoted
as \Zob, is particularly interesting. Antiquarks are also assumed in
the notation, and the $\PZ/\gamma^{\ast}$ interference contribution
is considered to be part of the process.  Within the SM, the \Zob
final state is the dominant background for studies of the associated
production of Higgs and \PZ bosons, in which the Higgs boson decays
into a \bbbar pair~\cite{Chatrchyan:2013zna}. Many physics scenarios
beyond the SM predict final states with $\PQb$ quarks and \PZ bosons:
new generations of heavy quarks ($\PQb^{\prime}, \PQt^{\prime}$)
decaying into \Zob~\cite{Holdom:2009rf}, supersymmetric Higgs bosons
produced in association with $\PQb$ quarks~\cite{Hall:2011aa}, and
extended SM scenarios with additional SU(2) doublets with enhanced
$\PZ\bbbar$ coupling~\cite{Choudhury:2001hs}. The study of the
associated production of \PZ bosons and $\PQb$ quark jets may also
provide information useful in describing an analogous process where a
$\PW$ boson is produced, which is more difficult to measure because of
higher background contamination.
\par}

This paper presents measurements of associated production of a \PZ
boson and b quark jets using proton-proton collision data at 8\TeV
collected with the CMS detector, corresponding to an integrated
luminosity of 19.8\fbinv.  The \PZ boson is reconstructed through its
leptonic decay into an electron or muon pair, while the presence of
$\PQb$ quarks is inferred from the characteristics of jets (denoted
as $\PQb$ jets) that originate from their hadronization products and
subsequent decays.  In order to characterize \Zob production, fiducial
differential cross sections are measured as a function of five
kinematic observables: the transverse momentum \pt and pseudorapidity
$\eta$ of the highest-\pt $\PQb$ jet, the \PZ boson \pt, the scalar
sum of the transverse momenta of all jets regardless of the flavour of
the parton producing them (\HT), and the azimuthal
angular difference between the direction of the \PZ boson and the
highest-\pt $\PQb$ jet ($\Delta \phi_{\PZ\PQb}$). Ratios of the
differential cross sections for \Zob and $\PZ$+jets production,
inclusive in jet flavour, are also measured as a function of these
five observables. The cancellation of several systematic uncertainties
in the cross section ratio allows an even more precise comparison with
theory than the differential cross sections themselves.

Events with at least two $\PQb$ jets, henceforth \Ztb, contribute as
background sources to other SM and beyond-SM processes.  The
production dynamics of this kind of event are studied through the
measurement of the fiducial differential cross section as a function
of observables characterizing the kinematic properties of the dijet
system formed by the leading and subleading (in \pt) $\PQb$ jets: the
\pt of these two jets; the \PZ boson \pt; the invariant masses of the
$\PQb\PQb$ and $\PZ\PQb\PQb$ systems ($M_{{\PQb\PQb}}$ and
$M_{\PZ\PQb\PQb}$ respectively); the angle $\Delta\phi_{{\PQb\PQb}}$
between the two $\PQb$ jets in the plane transverse to the beam axis
and their separation in the $\eta$-$\phi$ plane ($\Delta
R_{{\PQb\PQb}}$); the distance in the $\eta$-$\phi$ plane between the
\PZ boson and the closer $\PQb$ jet ($\Delta
R^{\text{min}}_{\PZ\PQb}$); and the asymmetry in the distances in
the $\eta$-$\phi$ plane between the \PZ boson and the closer versus
farther $\PQb$ jets ($A_{\PZ\PQb\PQb}$).

Previously, the cross section for the associated production of \PZ
bosons and $\PQb$ jets was measured in proton-antiproton collisions
by the CDF~\cite{Aaltonen:2008mt} and D0~\cite{Abazov:2010ix}
Collaborations at the Fermilab Tevatron and in proton-proton
collisions at a centre-of-mass energy of 7\TeV by the
ATLAS~\cite{Aad:2014dvb} and CMS~\cite{Chatrchyan:2013zja}
Collaborations at the CERN LHC. The CMS Collaboration also studied
\Ztb production by explicitly reconstructing $\PQb$ hadron
decays~\cite{Chatrchyan:2013zjb}, in order to explore the region where
$\PQb$ quarks are emitted in an almost collinear topology.  Previous
measurements of the ratio of the \Zob to the $\PZ$+jets inclusive cross
section were published by the D0
Collaboration~\cite{Abazov:2013uza}.

The paper is organized as follows: Section 2 is dedicated to the
description of the CMS apparatus and Section 3 to the data and
simulated samples used in the analysis. Section 4 discusses the
lepton, jet, and $\PQb$ jet reconstruction and the event
selection. Section 5 discusses background estimation, while Section 6
is dedicated to the description of the unfolding procedure to correct
data for detector effects. Section 7 presents a discussion of the
systematic uncertainties. In Section 8 the measured
differential cross sections and the corresponding ratios are
presented, together with a discussion of the comparison with
theoretical predictions. Finally, the results are summarized in
Section 9.

\section{The CMS detector}
A detailed description of the CMS detector, together with the
definition of the coordinate system used and the relevant kinematic
variables, can be found in Ref.~\cite{Chatrchyan:2008zzk}.  The
central feature of the CMS apparatus is a superconducting solenoid of
6\unit{m} internal diameter. The field volume houses a silicon
tracker, a crystal electromagnetic calorimeter (ECAL), and a brass and
scintillator hadron calorimeter, each composed of a barrel and two
endcap sections.  The magnet flux-return yoke is instrumented with
muon detectors.  The silicon tracker measures charged particles within
the pseudorapidity range $\abs{\eta} < 2.5$. It consists of 1440
silicon pixel and 15\,148 silicon strip detector modules and is
located in the 3.8\unit{T} field of the superconducting solenoid. For
nonisolated particles of $1 < \pt < 10\GeV$ and $\abs{\eta} < 1.4$,
the track resolutions are typically 1.5\% in \pt and 25--90
(45--150)\mum in the transverse (longitudinal) impact parameter
\cite{TRK-11-001}.  The electron momentum is estimated by combining
the energy measurement in the ECAL with the momentum measurement in
the tracker. The momentum resolution for electrons with $\pt
\approx 45\GeV$ from $\PZ \to \Pe \Pe$ decays ranges from
1.7\% for nonshowering electrons in the barrel region to 4.5\% for
showering electrons in the endcaps~\cite{Khachatryan:2015hwa}.  Muons
are measured in the pseudorapidity range $\abs{\eta} < 2.4$, with
detection planes made using three technologies: drift tubes, cathode
strip chambers, and resistive plate chambers. Matching muons to tracks
measured in the silicon tracker results in a relative transverse
momentum resolution for muons with $20 <\pt < 100\GeV$ of 1.3--2.0\%
in the barrel and better than 6\% in the endcaps. The \pt resolution
in the barrel is better than 10\% for muons with \pt up to
1\TeV~\cite{Chatrchyan:2012xi}. Forward calorimeters extend the
pseudorapidity coverage provided by the barrel and endcap detectors.

The CMS detector uses a two-level trigger system. The first level of the
system, composed of custom hardware processors, uses information from the
calorimeters and muon detectors to select the most interesting events in a
fixed time interval of less than 4\mus. The high-level trigger processor
farm further decreases the event rate from around 100\unit{kHz} to less
than 1\unit{kHz} before data storage.

\section{Event simulation}
\label{sec:evsim}
{\tolerance=800
The associated production of a \PZ boson and jets is experimentally
reconstructed as two opposite-sign same-flavour electrons or muons
accompanied by jets and can be mimicked by various background sources:
\ttbar events, dibosons ($\PW\PW$, $\PW\PZ$, $\PZ\PZ$) and $\PW$
bosons produced in association with jets, single top quark events, as
well as $\PZ$+jets events in which the \PZ boson decays into
$\Pgt^+\Pgt^-$.  Diboson events with a leptonic \PZ boson decay and
jets produced in the hadronic decay of the other vector boson are not
considered as part of the signal.  Samples of simulated events are
used to model both the signal and the background processes.  The
\MADGRAPH 5.1.3.30~\cite{Alwall:2011uj} event generator is used to
simulate $\PZ$+jets (including jets from $\PQb$ quarks), \PW+jets, and
\ttbar events; this generator implements a leading-order (LO) matrix
element calculation with up to four (three) additional partons in the
final state for V+jets (\ttbar) events, using the CTEQ6L1 PDF
set~\cite{Pumplin:2002vw}, which is based on the five flavour scheme (5FS).
A detailed discussion is given in Section~\ref{sec:theory}.  The
parton-level events are interfaced with \PYTHIA version
6.424~\cite{Sjostrand:2006za} for parton showering, hadronization, and
description of the multiple-parton interactions (MPIs).  The \PYTHIAS
Z2* tune, which is based on the CTEQ6L1 PDF set, is used~\cite{Chatrchyan:2013gfi}. The matrix
element and parton shower calculations are matched using the \kt-MLM
algorithm~\cite{Alwall:2007fs}. The cross section inclusive in jet
multiplicity is rescaled to its next-to-next-to-leading-order (NNLO)
prediction, computed with \FEWZ 3.1~\cite{Gavin:2010az,Li:2012wna} for
the $\Z$+jets and \PW+jets processes, and with the calculation of
reference~\cite{Czakon:2013goa} for the \ttbar process. To study
systematic uncertainties, signal events are also generated using
\AMCATNLO~\cite{Alwall:2014hca} version 2.2.1, with
next-to-leading-order (NLO) matrix elements for zero, one, and two
additional partons merged with the {\sc FxFx}
algorithm~\cite{Frederix:2012ps}, interfaced with \PYTHIA version
8.205~\cite{Sjostrand:2007gs} for showering and hadronization. In this
case the NNPDF 3.0 NLO PDF set~\cite{Ball:2014uwa} is used. Depending
on the flavours included in the matrix element calculation of the
event or produced in the parton shower through gluon splitting, the
inclusive $\PZ$+jets sample can be divided into $\PZ$+b quark, c
quark, and light-flavour (u, d, s quark and gluon) jet subsamples. As
explained in Section~\ref{sec:unfolding}, the jet flavour
identification is based on the particle content of the final state.
\par}

Diboson events are simulated with \PYTHIAS, and the inclusive cross
section rescaled to the NLO prediction provided by
\MCFM~\cite{Campbell:2011bn}.  The single top quark contribution is
evaluated using \POWHEGBOX version
1.0~\cite{Frixione:2007vw,Nason:2004rx,Alioli:2010xd,Alioli:2009je,Re:2010bp}
interfaced with \PYTHIAS for parton showering, hadronization, and MPI
description.  The contribution of multijet events is evaluated using
\PYTHIAS generated events and found to be negligible.

Generated events are processed with a simulation of the CMS
detector based on the \GEANTfour
toolkit~\cite{Agostinelli:2002hh}. Signals induced by additional
$\Pp\Pp$ interactions in the same or adjacent bunch crossings,
referred to as pileup, are simulated using events generated with
\PYTHIAS. The pileup distribution in simulation is adjusted in order
to reproduce the collision rates observed in data.  During the 2012
data taking, the average pileup rate was about 21 interactions per
bunch crossing.

\section{Event selection}
The analysis is based on an online trigger selection requiring events
to contain a pair of electron or muon candidates with asymmetric
minimum thresholds on their transverse momenta. These threshold
settings depended on the instantaneous luminosity and reached maximum
values of 17\GeV for the leading lepton and 8\GeV for the subleading
one.  Events are required to contain a \PZ boson, reconstructed through
its decay into an electron or muon pair, produced in association with
at least one or at least two hadronic jets. For the \Zob and \Ztb
event selections the jets are also required to be identified as
originating from the hadronization of a $\PQb$ quark.

All the measured particles are
reconstructed using the particle-flow (PF)
algorithm~\cite{CMS-PAS-PFT-09-001,CMS-PAS-PFT-10-001}.
The particle-flow event algorithm reconstructs and identifies each
individual particle with an optimized combination of information from
the various elements of the CMS detector. The energy of photons is
obtained directly from the ECAL measurement, corrected for
zero-suppression effects. The energy of electrons is evaluated from a
combination of the electron momentum at the primary interaction vertex
as determined by the tracker, the energy of the corresponding ECAL
cluster, and the energy sum of all bremsstrahlung photons spatially
compatible with originating from the electron track. The transverse momentum of
the muons is obtained from the curvature of the corresponding track. The
energy of charged hadrons is determined from a combination of the
momentum measured in the tracker and the matching ECAL and HCAL energy
deposits, corrected for zero-suppression effects and for the response
functions of the calorimeters to hadronic showers. Finally, the energy
of neutral hadrons is obtained from the corresponding corrected ECAL
and HCAL energies.

The reconstructed leptons selected as candidate
decay products of the \PZ boson must match those that triggered
the event and must be associated with the primary vertex of the event,
defined as the reconstructed vertex with the largest sum of $\pt^2$ of its
constituent tracks.
Reconstructed electrons must satisfy a set of selection criteria
designed to minimize misidentification at a desired efficiency
level~\cite{Khachatryan:2015hwa}; the discriminating observables
include the measured shower shape in the ECAL and the spatial
matching between the electromagnetic deposit in the calorimeter and
the reconstructed track associated with it. Additional requirements on
electron tracks are used to reject products of photon
conversions. Electron isolation criteria exploit the full
PF-based event reconstruction, using particles within a
cone around the electron direction with radius
$\DR = \sqrt{\smash[b]{(\Delta\phi)^2 + (\Delta\eta)^2}} = 0.3$.
The isolation requirement is defined by $I_{\text {rel}}=(I_{\text
{charged}} + I_{\text {photon}}+I_{\text {neutral}})/\pt^{\Pe} <
0.15$, where $I_{\text {charged}}$ is the scalar \pt sum of all the charged
hadrons, $I_{\text {photon}}$ is the scalar \pt sum of all the photons, and
$I_{\text {neutral}}$ the scalar sum of \pt of all the neutral hadrons in the
cone of interest. The notation $\pt^{\Pe}$ refers to the transverse momentum of the
reconstructed electron. Pileup can add extra particles, which affect the
isolation variable. Accordingly, only charged
particles originating from the reconstructed primary vertex are used in
the calculation of $I_{\text {charged}}$.  The photon and neutral
hadronic contribution to the isolation variable coming from pileup is
subtracted using the jet area approach~\cite{Cacciari:2007fd}.
Electrons must have $\pt^{\Pe} > 20\GeV$ and be reconstructed within the
pseudorapidity range $\abs{\eta}<1.44$ and $1.57<\abs{\eta}<2.4$,
which exclude the barrel-endcap transition region.

Muon identification criteria are based on the fit quality for tracks
measured in the tracker and the muon detector~\cite{Chatrchyan:2012xi}.
Further selection criteria are
added in order to reject muons from cosmic rays.  Muon isolation is
computed using all particles reconstructed by the PF
algorithm within a cone of radius $\DR = 0.4$ around the muon direction, requiring $I_{\text
{rel}}=(I_{\text {charged}}+I_{\text {photon}}+I_{\text
{neutral}})/\pt^{\Pgm} < 0.2$.  Muons must have $\pt^{\Pgm} >
20\GeV$ and $\abs{\eta}<2.4$. As in the case of electrons, charged particles
not originating from the primary vertex are excluded from the isolation calculation.
The pileup contribution to $I_{\text {photon}}$ and $I_{\text{neutral}}$
is estimated as half of the corresponding charged hadronic component and is
subtracted in the definition of the $I_{\text {rel}}$ variable.

The efficiencies for lepton trigger, reconstruction, identification,
and isolation are measured with the ``tag-and-probe"
technique~\cite{Khachatryan:2010xn} as a function of the lepton $\eta$
and \pt. A sample of events containing a \PZ boson decaying into
$\Pe^+\Pe^-$ or $\mu^+\mu^-$ is used for these studies.  Efficiency
corrections (``scale factors'') of up to 1.2\% (7.3\%), dependent on
lepton \pt and $\eta$, are applied to account for differences in the
estimated efficiencies between data and simulation in the electron
(muon) channel.

The pair of selected same-flavour, opposite-sign, highest-\pt isolated
leptons is retained as a \PZ boson candidate if the invariant mass
$M_{\ell\ell}$ of the pair lies within the 71--111\GeV mass interval.
The overall efficiency of the trigger and event selection within the
fiducial acceptance is 88\% for dimuons and 58\% for dielectrons.

Jets are reconstructed using the anti-\kt
algorithm~\cite{Cacciari:2008gp, Cacciari:2011ma} with a distance
parameter of 0.5. In order to suppress the contribution from pileup
interactions, charged particles not associated with the primary
vertex are excluded from the clustering. Jets are
required to be in the tracking acceptance region $\abs{\eta}<2.4$ and
to have $\pt > 30\GeV$, thereby reducing the contribution from the
underlying event to less than 5\%, where jets have a softer \pt spectrum compared to
jets from the hard scattering process. Jets with a distance $\DR <
0.5$ from the closer lepton used for the \PZ boson decay
reconstruction are not considered in the analysis.  The jet energy
scale (JES) is calibrated using a factorized approach as described in
Refs.~\cite{Chatrchyan:1369486,Khachatryan:2198719}. The jet energy resolution (JER) in
data is known to be worse than in the simulation; therefore the
simulated resolution is degraded to compensate for this effect as a
function of the jet kinematics~\cite{Chatrchyan:1369486,Khachatryan:2198719}.

Jets from $\PQb$ quarks are identified using the combined secondary
vertex (CSV) b tagging algorithm~\cite{Chatrchyan:2012jua}, a
multivariate classifier that makes use of information about
reconstructed secondary vertices as well as the impact parameters of
the associated tracks with respect to the primary vertex to
discriminate \PQb jets from $\PQc$ and light-flavour jets. The
threshold applied to the discriminating variable gives a b tagging
efficiency of about 50\% and a misidentification probability of 0.1\%
for light jets and 1\% for $\PQc$ jets. Scale factors, measured in
multijet events and dependent on jet \pt, are used to correct the b,
c, and light-flavour jet efficiencies in the simulation to match those
observed in the data~\cite{Chatrchyan:2012jua}.  The scale factors for
$\PQb$ jets are determined using samples of events enriched in such a
flavour of jets. This enrichment is obtained including both multijet
events containing a muon geometrically associated with a jet, with
high probability of originating from the semileptonic decay of a
$\PQb$ hadron, and leptonic and semileptonic \ttbar events, where the
leading \pt jets are usually $\PQb$ jets. The scale factors are around
0.93, slowly decreasing for jets with \pt above 120\GeV.  The scale
factors for $\PQc$ jets are assumed the same as for $\PQb$ jets, with
an uncertainty twice as large. Relatively pure samples of $\PQc$ jets
from $\PW+\PQc$ events, selected using identified muons within the
jet, are used to validate this assumption.  For light-flavour jets,
the same CSV algorithm yields scale factors between 1.1 and 1.4,
depending on the jet \pt.  The calculation is based on tracks with
negative signed impact parameter and secondary vertices with negative
signed decay lengths, where the sign is defined by the relative
direction of the jet and the particle momentum.  Finally, events are
selected if they contain a \PZ boson candidate and at least one
$\PQb$-tagged jet.

The missing transverse momentum vector \ptvecmiss is defined as the
projection on the plane perpendicular to the beams of the negative
vector sum of the momenta of all reconstructed particles in an
event. Its magnitude is referred to as \ETmiss.  The \ETmiss
significance, introduced in
Ref.~\cite{2011JInst...6.9001C,Khachatryan:2014gga}, offers an
event-by-event assessment of the consistency of the observed missing
energy with zero, given the reconstructed content of the event and
known measurement resolutions.  In order to suppress the background
contamination from \ttbar production, events with \ETmiss significance
greater than 30 are vetoed. This requirement provides a 13\%
\ttbar background rejection with no loss in signal efficiency.

\begin{figure*}[hbtp]
\begin{center}
\includegraphics[width=\ghmFigWidth]{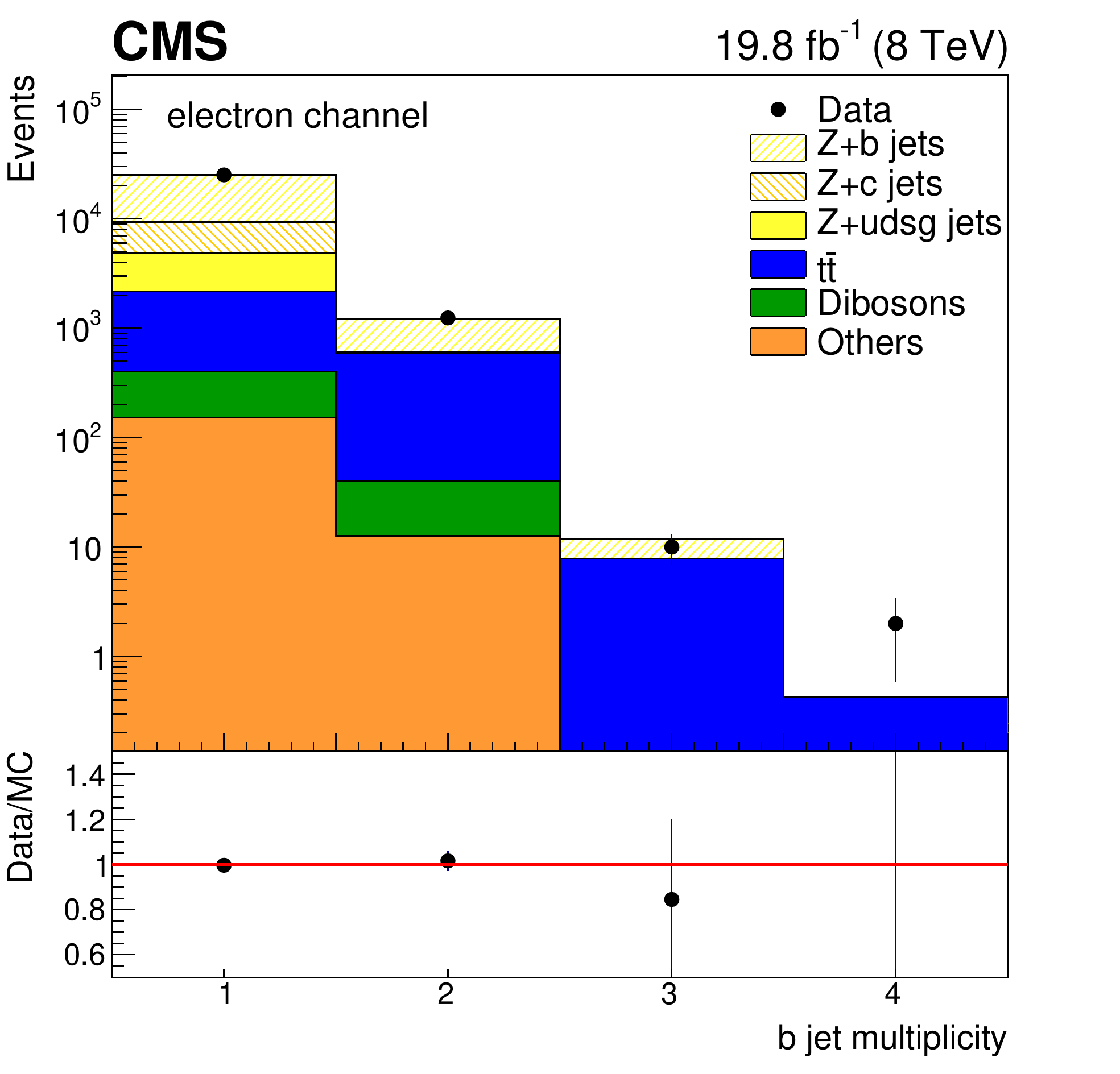}
\includegraphics[width=\ghmFigWidth]{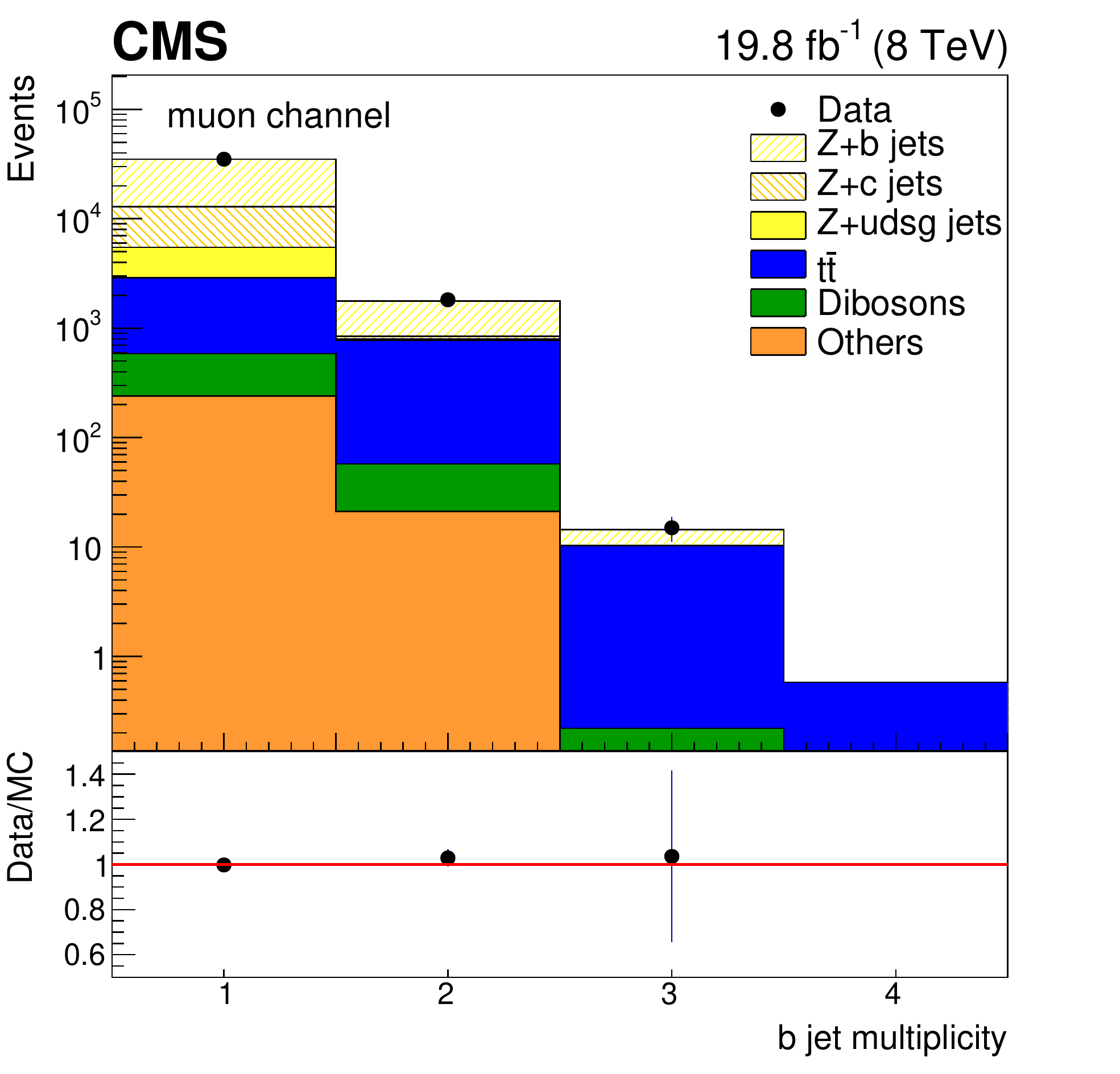}
\end{center}
\caption{Exclusive $\PQb$-tagged jet multiplicity distributions for \Zob
events, for the electron (left) and muon (right) decay channel of \PZ
boson. Error bars account for statistical uncertainties in data
in the upper plots and in both data and simulation in the bottom
ratio plots, that show the data to MC ratio.}
\label{fig:multb}
\end{figure*}

\begin{figure*}[hbtp]
\begin{center}
\includegraphics[width=\ghmFigWidth]{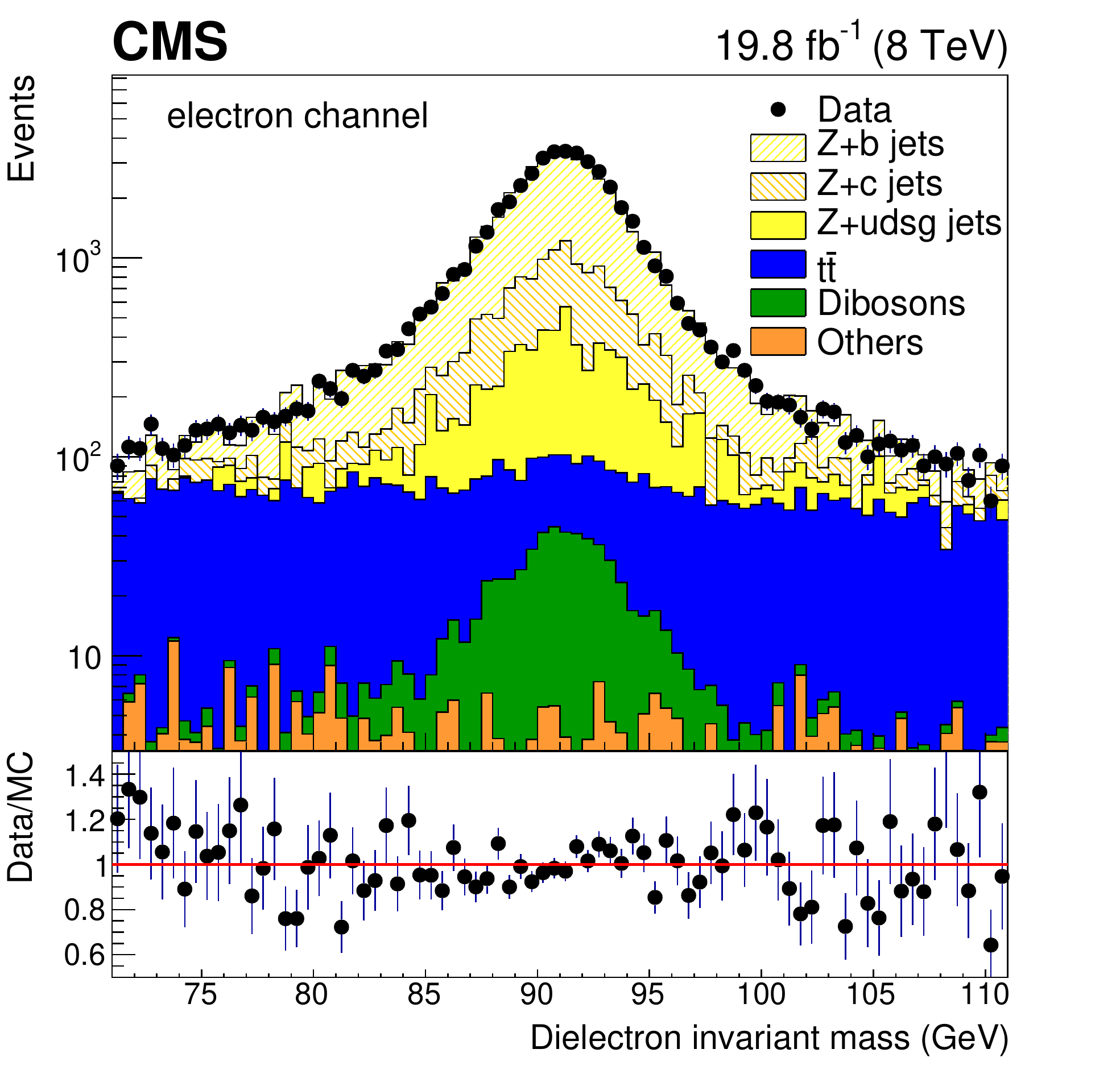}
\includegraphics[width=\ghmFigWidth]{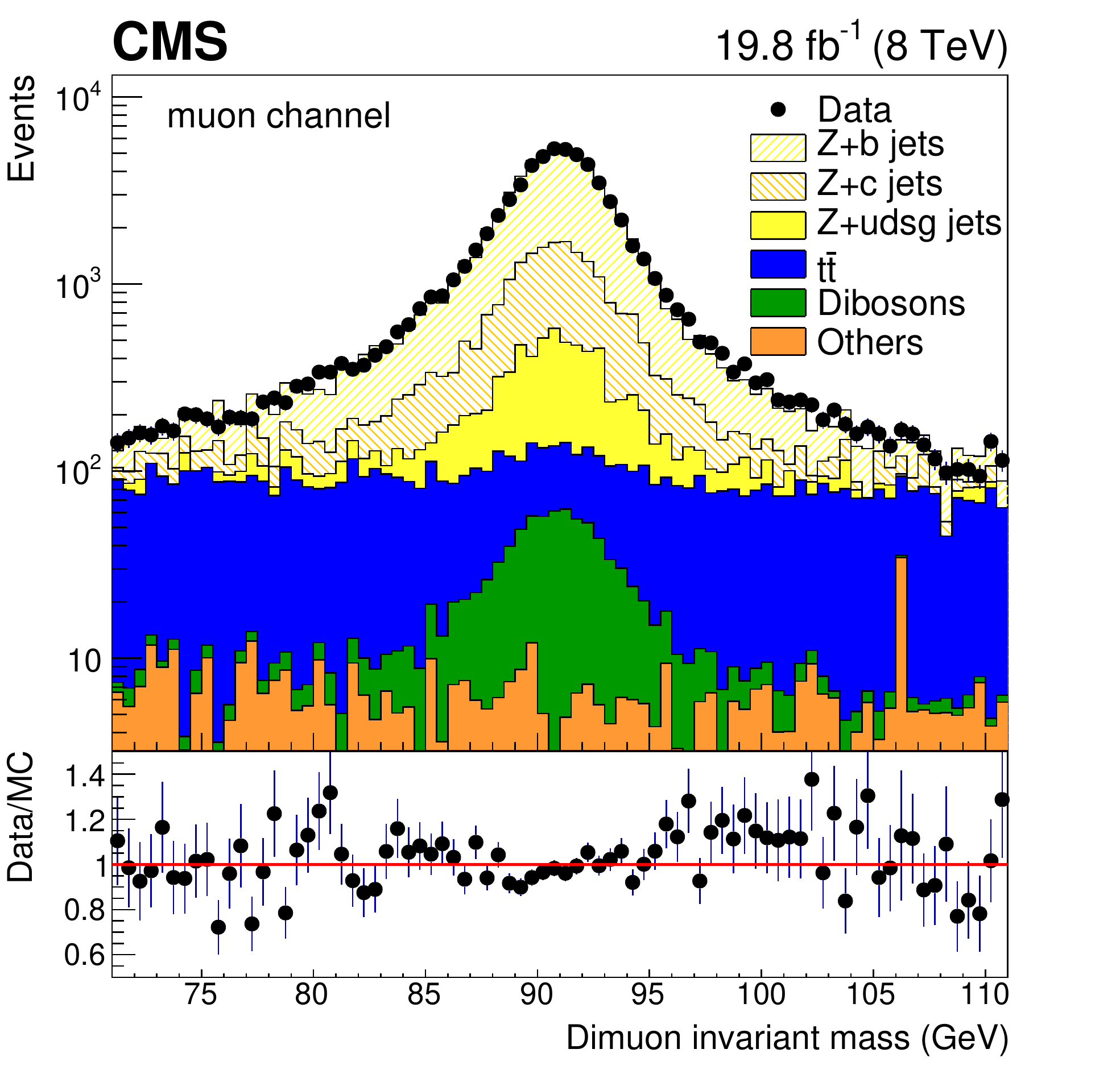}
\end{center}
\caption{Dilepton invariant mass distributions for \Zob events, for
the electron (left) and muon (right) \PZ boson decay channels. Error
bars account for statistical uncertainties in data in the upper
plots and in both data and simulation in the bottom ratio plots, that
show the data to MC ratio.}
\label{fig:invzb}
\end{figure*}

The \Zob event selection described above yields 26443 (36843) events
for the dielectron (dimuon) channels.  The exclusive $\PQb$-tagged jet
multiplicity and invariant mass distributions of the same flavour
dilepton are presented in Figs.~\ref{fig:multb} and \ref{fig:invzb},
for the \Zob event selection for electron and muon respectively. Data
are compared with the simulations where the $\Z$+jets events are
described by \MADGRAPH+\PYTHIAS, and good agreement is observed. In
all figures, the simulated events are reweighted by scale factors in
order to compensate for the residual data-to-simulation discrepancies
in lepton selection efficiency, JES and JER calibration, and b tagging
efficiency. The background contributions from $\Z$+jets and \ttbar
events as adjusted in Section~\ref{sec:backg} are included in
Figs.~\ref{fig:multb} and~\ref{fig:invzb}.

\section{Background estimation}
\label{sec:backg}
A Drell--Yan event in which a \PZ boson decays into $\tau^+\tau^-$ may
contribute to the dielectron or dimuon signal events if both $\Pgt$
leptons decay into electrons or muons. These events are treated as a
background source and, being at the few per mil level, their
contribution is evaluated from simulation.

The process $\Pp\Pp\to \ttbar \to \PW^+ \PQb \PW^-{\PAQb}\to
\ell^+\ell^-\PQb{\PAQb}+\MET$ is the dominant non-Drell--Yan
background source. The \ttbar background contribution is estimated
separately for $\PZ$+jets, \Zob, and \Ztb events by using the signal
selection criteria to produce samples of $\Pe\Pgm$ pairs, which are
enriched in \ttbar events with negligible signal contamination.  For
each measured observable these samples provide the estimates of the
\ttbar background; residual non-\ttbar backgrounds in them, amounting to about
29\%, 8\% and 2\% respectively, are subtracted using the simulated
prediction.  The integrals of such estimates need to be rescaled by
the ratio of the same-flavour lepton to $\Pe\Pgm$ yields. This ratio
is determined using control samples for both the same-flavour lepton
and $\Pe\Pgm$ selections by inverting the \MET significance
requirement, namely, \MET significance $>$30.  For the same-flavour
lepton samples, this selection removes the contribution from the
signal processes, while enhancing the fraction of \ttbar events in the
sample. The residual contamination from other non-\ttbar processes is
similar in the same-lepton and $\Pe\Pgm$ selections, amounting to
about 20\%, 7\%, 3\% respectively, and is again taken into account
using the simulation.  The ratio of the $\Pe\Pgm$ to the $\Pe\Pe$ or
$\Pgm\Pgm$ yields in the control samples is used to rescale the
estimates of this background source for each lepton channel
separately. The ratio is determined as the scaling factor for the
normalization of the binned dilepton invariant mass ($M_{\ell\ell}$)
distribution in the $\Pe\Pgm$ sample that minimizes the difference of
this distribution from the corresponding same-lepton-flavour
$M_{\ell\ell}$ distribution in a least-square fit procedure. The fit
of the $M_{\ell\ell}$ distribution is performed in the sideband
regions 50--84\GeV and 100--200\GeV, to avoid any assumption about the
$M_{\ell\ell}$ shape for both opposite and same-sign lepton pairs in
the \PZ peak region.

The remaining background sources are estimated using simulation. Diboson
events may mimic the $\PZ$+b final state when one or more leptons are not
reconstructed or when a $\PW$ or \PZ boson decays hadronically into a
$\cPq\cPaq$ pair (in particular a \PZ boson may decay into a genuine
\bbbar pair). Single top quarks produced in association with either a
$\PW$ boson or one or more $\PQb$ jets may also generate a signal-like
signature.  These events, together with $\PW$+jets, can mimic the
signal if a lepton of the same flavour is produced in the
hadronization or if a hadron is misidentified. The contribution of
multijet events is found to be negligible, as has been previously
observed in other similar $\PZ$+jets analyses~\cite{azimuthalref}.

After subtraction of all non-Drell--Yan background contributions, the
extraction of the \Zob and \Ztb event yields requires an evaluation of
the purity of the b tagging selection, i.e.\ the fraction of selected
Drell--Yan events in which the desired number of $\PQb$-tagged jets,
at least one or at least two, originate from the hadronization of a
corresponding number of $\PQb$ quarks. This fraction is determined
from a study of the secondary vertex mass distribution of the leading
$\PQb$-tagged jet, defined as the invariant mass of all the charged
particles associated with its secondary vertices, assuming the pion
mass for each considered particle. This evaluation is done separately
for dielectron and dimuon final states to avoid correlations between
channels and to simplify the combination. The secondary vertex mass
distributions for $\PQb$, $\PQc$, and light-flavour jets produced in
association with \PZ bosons are obtained from the simulation based on
the \MADGRAPH event generator interfaced with \PYTHIAS by using the
5FS scheme for PDFs. The sum of the distributions is fitted to
the observed distribution with an extended binned likelihood, after
subtraction of all non-Drell--Yan background contributions, by varying
the three normalization scale factors $c_{\mathrm{b}}$,
$c_{\mathrm{c}}$, $c_{\mathrm{udsg}}$ for the various components. The
$c_{\mathrm{c}}$, $c_{\mathrm{udsg}}$ factors are used for the
subtraction of the respective components. This procedure reduces the
dependence on the normalization of the $\cPqb$ hadron production and decay
in the simulation because the expected shape of the secondary vertex
mass distribution is used. In the case of the \Ztb selection, as it
can be seen in Fig.~\ref{fig:multb}, the contamination from c and
light-flavour jets is negligible and is subtracted using simulation;
only the $c_{{\PQb\PQb}}$ scaling factor for the genuine double $\PQb$
jet component is determined from the fit, and it is used only to
correct the relative proportion of \Zob and \Ztb events in the
simulation, as discussed in Section~\ref{sec:unfolding}.

\begin{figure*}[tb]
\begin{center}
\includegraphics[width=\ghmFigWidth]{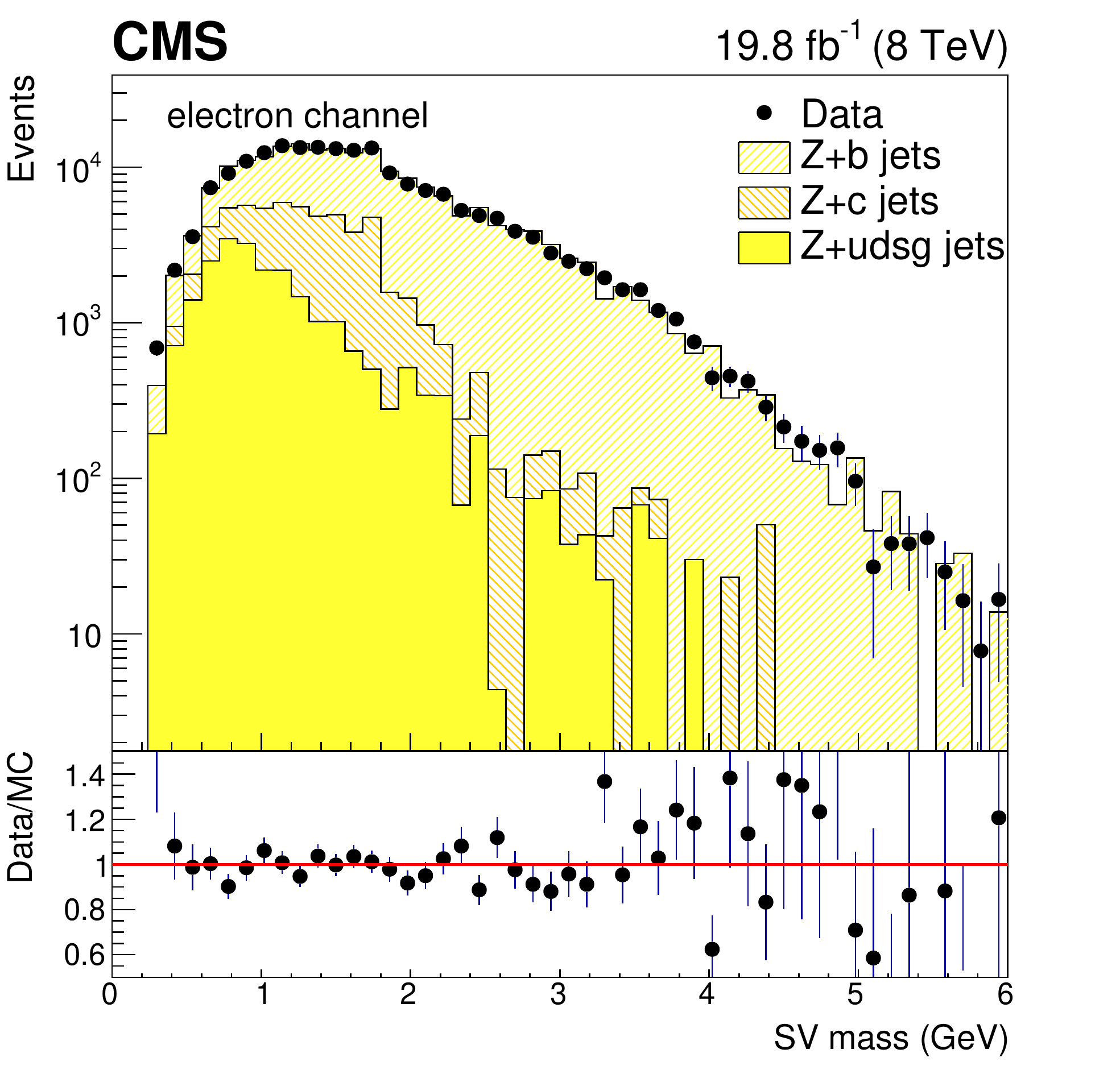}
\includegraphics[width=\ghmFigWidth]{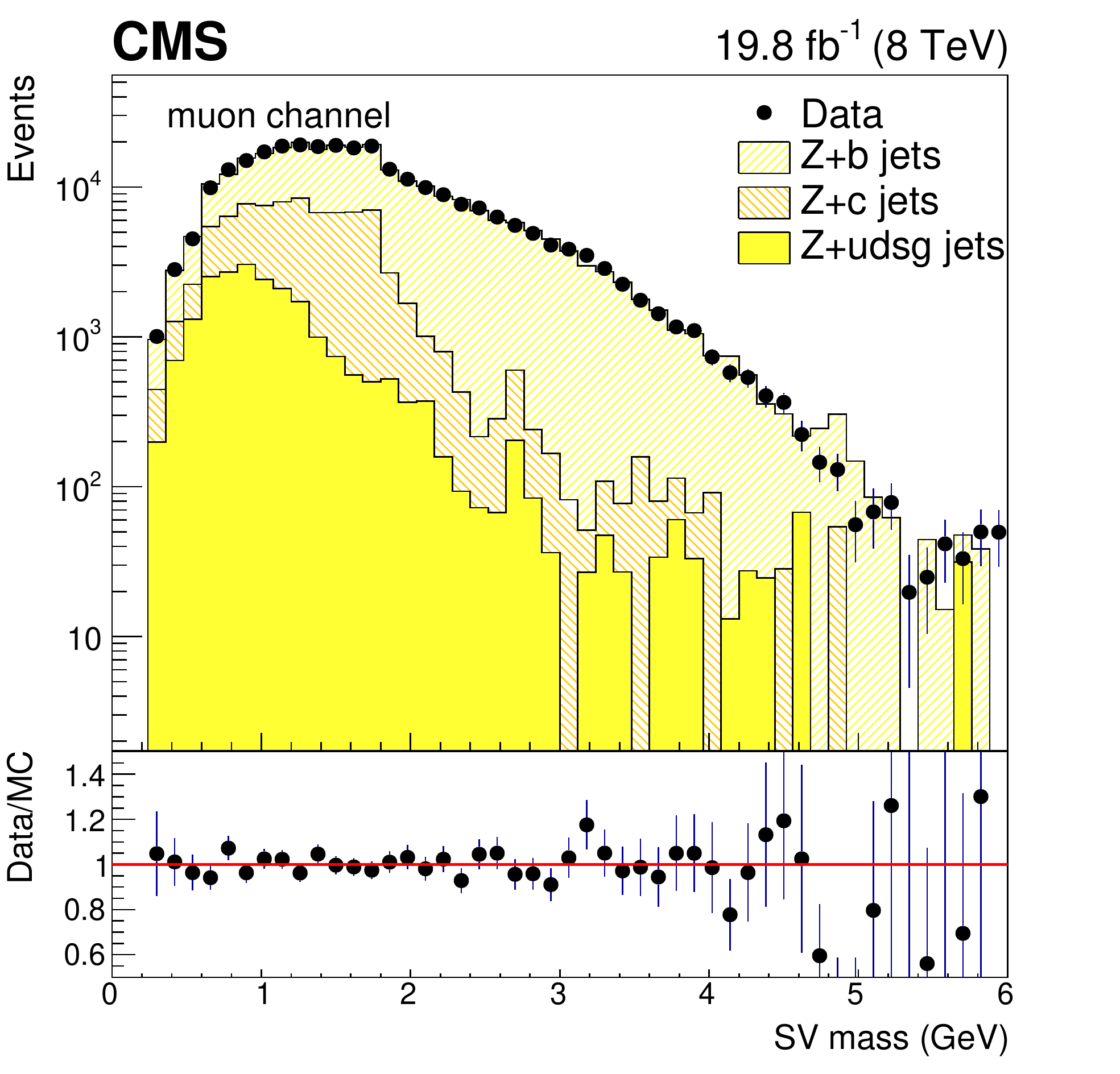}
\end{center}
\caption{Distributions of the secondary vertex (SV) mass of the
leading jet after the \Zob selection with the \PZ boson decaying into
electrons (left) and muons (right). The subsamples corresponding to
$\PQb$-tagged jets originating from $\PQb$, c, and light-flavour
quarks or gluons are shown, with normalizations determined in the
fit to data. Non-Drell--Yan background sources are subtracted. Error
bars account for statistical uncertainties in
data in the upper plots and in both data and simulation in the
bottom ratio plots.
\label{fig:bpurity1b} }
\end{figure*}

The results of the fit to the secondary vertex mass distributions are
presented in Fig.~\ref{fig:bpurity1b} for the \Zob analysis, showing
the flavour composition in each channel. Data-to-simulation scale
factors, as determined by the fit, are given in
Table~\ref{tab:Fraction-of-beauty} for both event selections and \PZ
boson decay channels. The flavour composition of the selected sample
after the scale factor corrections for the \Zob samples is
also shown.

The $\PQb$-flavour contribution is constrained by the high secondary vertex
mass region of the distribution of the CSV discriminating variable,
while the $\PQc$-flavour contribution is mostly important in the region
between 1 and 2\GeV, and the light-flavour contribution below
1\GeV. This results in a strong anticorrelation both between the $\PQb$- and
$\PQc$-flavour and between $\PQc$- and light-flavour contributions,
with an estimated correlation coefficient of about -0.6 in both cases,
whereas the correlation between the $\PQb$- and light-flavour
contributions is negligible.  As a consequence, a fluctuation in the
small $\PQc$ quark component may cause a difference in the scale
factors between different lepton channels.
\begin{table*}[htb]
\topcaption{Normalization scale factors and post-fit fractions for b,
  c and light-flavour (u, d, s quark and gluon) components in the
  selected \Zob events, and scale factor for b in the selected \Ztb events,
  obtained from a fit to the secondary vertex mass distribution for
  dielectron and dimuon final states. The quoted uncertainties are
  statistical only.}
\centering
\begin{tabular}{lcccrrr}
Event selection & \emph{$c_{\mathrm{b}}$} & \emph{$c_{\mathrm{c}}$} & \emph{$c_{\mathrm{udsg}}$} & \Zob (\%) & Z+c (\%) & Z+udsg (\%) \\
\hline
\Zob ($\Pe\Pe$)   & $0.91\pm0.02$ & $1.29\pm0.13$ & $1.70\pm0.21$ &
$69.5\pm1.8$ & $19.0\pm2.0$ & $11.4\pm1.4$ \\
\Zob ($\Pgm\Pgm$) & $0.91\pm0.02$ & $1.51\pm0.12$ & $1.18\pm0.19$ &
$69.7\pm1.5$ & $22.4\pm1.8$ & $7.9\pm1.2$ \\
\end{tabular} \\
\begin{tabular}{lc}
Event selection & \emph{$c_{{\PQb\PQb}}$} \\
\hline
\Ztb ($\Pe\Pe$)   & $1.18\pm0.12$ \\
\Ztb ($\Pgm\Pgm$) & $1.17\pm0.09$ \\
\end{tabular}
\label{tab:Fraction-of-beauty}
\end{table*}

The signal yield for \Zob events is therefore obtained, for each bin
of a distribution, from the selected event yield $N^{\text{selected}}$ as
\begin{equation*}
\ifthenelse{\boolean{cms@external}}
{
\begin{split}
N_{\Zob}  =
N^{\text{selected}}_{\Zob} & - N_{\ttbar}
  - N_{\text{Dibosons}}^{\mathrm{MC}} - N_{\text{Others}}^{\mathrm{MC}} \\
   & -  c_{\mathrm{c}} N_{\mathrm{Z+c}}^{\mathrm{MC}} -
c_{\mathrm{udsg}} N_{\mathrm{Z+udsg}}^{\mathrm{MC}} ,
\end{split}
}
{
{N_{\Zob} =
N^{\text{selected}}_{\Zob} - N_{\ttbar} - N_{\text{Dibosons}}^{\mathrm{MC}} - N_{\text{Others}}^{\mathrm{MC}} - c_{\mathrm{c}} N_{\mathrm{Z+c}}^{\mathrm{MC}} -
c_{\mathrm{udsg}} N_{\mathrm{Z+udsg}}^{\mathrm{MC}} ,}
}
\end{equation*}
where $N_{\ttbar}$, $N_{\text{Dibosons}}^{\mathrm{MC}}$, and $N_{\text{Others}}^{\mathrm{MC}}$ are
the \ttbar, diboson, and other background contributions respectively,
$c_{\mathrm{c}} N_{\mathrm{Z+c}}^{\mathrm{MC}}$ and $c_{\mathrm{udsg}}
N_{\mathrm{Z+udsg}}$ are the numbers of Drell--Yan events in which the b-tagged
jets originate from either a c or a light-flavour parton, and the scale factors
multiply the event yields predicted by the simulation. For the calculation of the
\Ztb event yield a similar procedure is applied:
\begin{equation*}
{N_{\Ztb} =
N^{\text{selected}}_{\Ztb} - N_{\ttbar} -
N_{\text{Dibosons}}^{\mathrm{MC}}
- N_{\text{Others}}^{\mathrm{MC}} .}
\end{equation*}
The $c_\mathrm{c}$ and $c_\mathrm{udsg}$ scale factors are also
re-evaluated from subsamples obtained by dividing the ranges of the
studied observables into wide intervals, in order to study a possible
correlation with the observables themselves. The statistical
uncertainty of these scale factors depends on the chosen observable
and binning, ranging from a factor of 2 up to 10 relative to the size of the
uncertainty of the default values obtained with the full sample.
Because no statistically significant dependence is observed, the scale
factors estimated from the overall sample are used.

The amount of background in the final event selection, estimated with
the procedures discussed above, can be observed in
Fig.~\ref{fig:multb}. For the \Zob selection, in the electrons (muons)
samples the Z+$\PQc$ contribution amounts to about 17\% (20\%), the
Z+light flavour jets (including gluons) to 10\% (7\%), and the \ttbar
to 9\% (8\%). Other background contributions are globally below the
2\% level. The \Zob contribution in the corresponding
selected sample is about 62\% (63\%) for the electrons (muons) channel.

\section{Unfolding}
\label{sec:unfolding}
The differential event yields are corrected for event
selection efficiencies and for detector resolution effects back to the
stable-particle level. For this purpose, the singular value
decomposition (SVD)~\cite{Hocker:1995kb} unfolding technique,
implemented in the {\sc RooUnfold} toolkit~\cite{2011arXiv1105.1160A},
is used. The unfolding procedure is based on a response matrix, which
describes the relationship between the particle levels and measured values of a
given observable due to the detector resolution and acceptance.  The
response matrix is calculated using \Zob events that are generated with
\MADGRAPH in the 5FS, interfaced to \PYTHIAS, and followed by the
detector simulation. Response matrices are computed separately for
the \Zob and \Ztb selections. The proportion of events with
exactly one or at least two b quarks in the simulation is reweighted
to match that observed in data, as determined by the $c_{{\PQb\PQb}}$
scaling factor.

Fiducial cross sections are defined, based on event generator
predictions at the particle level, for leptons and jets reconstructed
from the collection of all stable final-state particles, using the
same selection criteria as the data analysis.  The two leptons
(electrons or muons) with the highest transverse momentum and with
$\pt > 20\GeV$ and $\abs{\eta} < 2.4$ are selected as \PZ boson decay
products if their invariant mass is in the range of 71--111\GeV.
Electromagnetic final-state radiation effects are
taken into account in the generator-level lepton definition by clustering
all photons in a cone of radius $\DR = 0.1$ around the final-state
leptons. The leptons selected as \PZ boson decay products are then
removed from the particle collection used for the jet clustering at
the generator level. The remaining particles, excluding neutrinos, are
clustered into jets using the anti-\kt algorithm with a distance parameter of
$0.5$. Generated jets are selected if their distance from the
leptons forming the \PZ boson candidate is larger than $\DR = 0.5$.
Jets originating from the hadronization of $\PQb$ quarks are selected
if a $\PQb$ hadron is an ancestor of one of the particles clustered
in it, and this $\PQb$ hadron has a distance from the jet in the
$\eta$-$\phi$ plane of $\DR \le 0.5$.  Jets and \PQb jets are selected if
they have $\pt > 30\GeV$ and lie in the pseudorapidity range
$\abs{\eta} < 2.4$.

As a cross-check of the SVD technique, the unfolding is also performed
with the iterative D'Agostini method~\cite{D'Agostini1995487},
leading to compatible results within the statistical uncertainties.

\section{Systematic uncertainties}
Several sources of systematic uncertainty affect the cross section
measurement: the JES and JER, the calculation of the unfolding
response matrix, the estimation of the $\PQb$ quark fraction, the
background subtraction, the event selection efficiencies, the pileup
description, and the integrated luminosity.  For every source other than
the luminosity, the full analysis procedure is repeated after the
variation of the corresponding input values, and the difference of the
extracted cross section with respect to the central measurement is
used as an estimate of the uncertainty due to that source. The
uncertainties are symmetrized, if not already symmetric. The
systematic uncertainties in the measured \Zob and \Ztb differential
cross sections are summarized in Table~\ref{tab:syst_Zb} and in
Tables~\ref{tab:syst_Zb_2bSample} and
\ref{tab:syst_Zb_2bSample_extra}, respectively.

Reconstructed jet energies must be corrected for several effects,
such as pileup contamination, instrumental noise, nonuniformities
and nonlinearities in the detector response, and flavour
composition.  The resulting uncertainty depends on the transverse
momentum and pseudorapidity of the jet. The systematic effect due
to the application of JES corrections in the data is estimated by
increasing and decreasing the correction parameters deviation from
their nominal values  by one standard. The uncertainty for the JER
correction is evaluated in the same way.

{\tolerance=800
For the cross section measurement in a given bin, the systematic
uncertainty induced by the model used in the unfolding procedure is
evaluated as the difference between the standard result and that
obtained with an alternative model for unfolding, namely \AMCATNLO
interfaced with \PYTHIAE. This alternative model implements NLO hard
scattering matrix elements, compared to the LO matrix elements of
\MADGRAPH interfaced to \PYTHIAS, and also includes different details of the
underlying event, hadronization, and particle decay descriptions
compared to the default choice.  In order to evaluate the
genuine model-induced effects, the statistical uncertainties from the
two simulated samples are subtracted in quadrature from the
difference; any negative results so obtained are replaced with
zero. The uncertainty associated with the size of the simulated sample
used to compute the response matrix elements is determined by
producing replicas of the matrix whose elements are
fluctuated according to a Poisson distribution.
\par}

The uncertainty induced by the secondary vertex mass fit, used to
extract the true flavour composition of the \Zob sample, is
twofold. One part is due to the statistical uncertainty in the
$c_\mathrm{c}$, $c_\mathrm{udsg}$ scale factors, whose effect is
estimated by varying them up and down by one standard deviation,
taking into account their correlation. This source of uncertainty is
considered as part of the statistical uncertainty, because it is due
to the finite size of the collision data sample. The other part stems
from the choice of the simulation model for the shape of the secondary
vertex mass distributions. This choice affects also the correction of
the relative proportion of different $\PQb$ multiplicities provided by
the scale factor $c_{\PQb\PQb}$. In addition, a systematic uncertainty
is associated, for both \Zob and \Ztb samples, with the modelling of
the $\PQc$ quark and light-flavour contributions to each measured
observable. Both of these model-induced uncertainties, collectively
indicated in the tables as ``c, udsg background model'', are estimated
by replacing the default model given by \MADGRAPH 5FS interfaced with
\PYTHIAS with \AMCATNLO 5FS interfaced with \PYTHIAE. The scale
factors, which are determined from the alternative model, are in
statistical agreement for dielectron and dimuon channels within one
standard deviation. The difference between the results obtained with
the two models is therefore considered as safely accounting for
possible residual discrepancies between data and simulation.

For each lepton channel the systematic uncertainties in the lepton
efficiency calculations for triggering, reconstruction,
identification, and isolation are estimated from the $\PZ\to\ell\ell$
``tag-and-probe'' measurements of data-to-simulation efficiency scale
factors. The global effect of the systematic uncertainty related to
the scale factors is 1.5\% in the dielectron final state and 2\% in
the dimuon final state. The uncertainties in the b tagging efficiency
scale factors include contributions from the pileup contamination, the
gluon splitting rate in simulation ($\cPg\to \bbbar$), varied by
${\pm}50\%$, and the energy fraction carried by the b hadrons in the
hadronization (varied by ${\pm}5\%$)~\cite{Chatrchyan:2012jua}.  The
global value of the $\PQb$ tagging systematic uncertainty amounts to
3\% per $\PQb$-tagged jet. Scale factors for $\PQc$ jets, assumed
equal to those for $\PQb$ jets, are assigned an uncertainty twice as
large as for the $\PQb$ jets.

The simulation is reweighted according to the generated primary
vertex multiplicity and the instantaneous luminosity in data to
reproduce the observed primary vertex multiplicity distribution,
and provide a reliable representation of pileup.
The minimum-bias event cross section in simulation is tuned to provide
the best agreement between data and simulation in the vertex multiplicity
distribution of $\PZ \to \Pgm \Pgm$ events. The uncertainty
associated with this procedure is estimated by varying this
minimum-bias cross section value by 5\%.

The uncertainty in the \ttbar background normalization is
derived from the statistical uncertainties of the same-flavour and
$\Pe\Pgm$ control samples and is included in the total
statistical uncertainty. The systematic uncertainty related to
the diboson background ($\PZ\PZ$, $\PW\PW$, $\PW\PZ$) is evaluated by
varying the theoretical production cross sections by ${\pm} 15\%$
of their central values, corresponding to about three standard
deviations of the overall theoretical normalization uncertainty and
covering the typical differences between the theoretical and measured values.
In addition, the statistical uncertainty induced by the
limited size of simulation samples is taken into account.

The systematic uncertainty in the integrated luminosity is
2.6\%~\cite{CMS-PAS-LUM-13-001}.

In the ratios of \Zob and \Ztb to the inclusive $\PZ$+jets cross
sections, the uncertainties
are simultaneously propagated to both the numerator and denominator,
taking correlations into account. The uncertainties in the energy scale,
resolution, and efficiency corrections for reconstructed leptons and jets
are considered to be fully correlated, as is the uncertainty in the
integrated luminosity.
Tables~\ref{tab:syst_Zb}--\ref{tab:syst_Zb_2bSample_extra}
summarize the ranges of variation of the uncertainties for each observable
measured with the \Zob and \Ztb samples.
\begin{table*}[htb]
\renewcommand{\arraystretch}{1.1}
\topcaption{Uncertainties (in percent) in the differential
cross sections as a function of the leading $\PQb$ jet \pt and
$\abs{\eta}$, the \PZ boson \pt, \HT, and $\Delta\phi_{\PZ\PQb}$ between the \PZ
boson and the leading $\PQb$ jet, for the \Zob sample.}
\centering
\begin{tabular}{lccccc}
Uncertainty (\%) & ${\rd\sigma}/{\rd\pt}$ & ${\rd\sigma}/{\rd\abs{\eta}}$ & ${\rd\sigma}/{\rd\pt^{{\PZ}}}$ & ${\rd\sigma}/{\rd H_{\mathrm{T}}}$ & ${\rd\sigma}/{\rd\Delta\phi_{\PZ\PQb}}$\\
\hline
JER                                & 0.3--1.7     & 0.1--0.6   & 0.2--2.6    & 0.4--1.9      & 0.1--2.2    \\
JES                                & 0.5--4.8     & 0.7--5.3   & 0.5--7.7    & 0.6--5.2      & 0.4--4.2    \\
Unfolding (MC model)               & 0.0--19.2    & 0.2--2.2   & 0.0--18.1   & 0.0--10.2     & 0.0--9.2    \\
Unfolding (MC statistics)          & 1.4--10.2    & 1.1--2.7   & 1.8--8.3    & 1.3--7.6      & 1.2--6.1    \\
c, udsg background model           & 0.0--6.1     & 0.0--7.0   & 0.0--19.9   & 0.7--7.5      & 0.0--10.9   \\
Electron (muon) efficiency         & 1.5 (2.0)    & 1.5 (2.0)  & 1.5 (2.0)   & 1.5 (2.0)     & 1.5 (2.0)   \\
$\PQb$ tagging efficiency         & 3.0          & 3.0        & 3.0         & 3.0           & 3.0         \\
Pileup                             & 0.2--4.3     & 0.6--1.4   & 0.4--2.0    & 0.2--2.3      & 0.2--1.6    \\
Background (systematic)            & 0.1--0.4     & 0.1--0.3   & 0.1--0.6    & 0.2--0.3      & 0.1--0.3    \\
Background (statistical)           & 1.2--7.2     & 1.0--2.5   & 1.5--5.8    & 1.3--4.6      & 1.2--5.9    \\
Integrated luminosity              & 2.6          & 2.6        & 2.6         & 2.6           & 2.6         \\
\hline
Total syst.\ uncertainty (\%)      & 5.5--21.7    & 5.2--10.6  & 5.6--22.8   & 8.4--13.8     & 6.0--13.3   \\
\hline
Total stat.\ uncertainty (\%)      & 2.6--8.8     & 3.0--5.4   & 2.9--8.6    & 3.1--6.0      & 3.1--7.0    \\
\end{tabular}
\label{tab:syst_Zb}
\end{table*}
\begin{table*}[htb]
\renewcommand{\arraystretch}{1.1}
\topcaption{Uncertainties (in percent) in the
differential cross sections as a function of the leading and
subleading $\PQb$ jet \pt, the \PZ boson \pt, the invariant mass of the
two $\PQb$-tagged jets, and the invariant mass of the \PZ boson and the
two $\PQb$-tagged jets, for the \Ztb sample.}
\centering
\ifthenelse{\boolean{cms@external}}{}{\resizebox{\textwidth}{!}}
{
\begin{tabular}{lccccc}
Uncertainty (\%) & ${\rd\sigma}/{\rd\pt^{\text{leading}}}$ & ${\rd\sigma}/{\rd\pt^{\text{subleading}}}$ & ${\rd\sigma}/{\rd\pt^{\PZ}}$ & ${\rd\sigma}/{\rd M_{{\PQb\PQb}}}$ & ${\rd\sigma}/{\rd M_{\PZ\PQb\PQb}}$\\
\hline
JER                                & 0.3--8.3     & 0.7--7.9    & 0.1--3.8    &  0.9--4.1    &  2.9--12.0   \\
JES                                & 4.4--17.0    & 7.7--23.3   & 3.1--20.3   &  6.7--15.3   &  3.8--16.2   \\
Unfolding (MC model)               & 0.0--74.4    & 0.0--52.6   & 0.0--53.6   &  0.0--37.8   &  0.0--57.3   \\
Unfolding (MC statistics)          & 8.0--39.4    & 9.0--35.8   & 8.8--27.0   &  7.6--28.0   &  10.0--20.8  \\
c, udsg background model           & 0.0--17.3    & 0.0--16.1   & 0.0--15.5   &  0.0--18.5   &  0.0--10.2   \\
Electron (muon) efficiency         & 1.5 (2.0)    & 1.5 (2.0)   & 1.5 (2.0)   &  1.5 (2.0)   &  1.5 (2.0)   \\
$\PQb$ tagging efficiency         & 6.0          & 6.0         & 6.0         &  6.0         &  6.0         \\
Pileup                             & 0.4--14.1    & 0.3--11.4   & 1.3--9.6    &  1.1--5.7    &  0.2--4.3    \\
Background (systematic)            & 0.3--0.9     & 0.1--0.7    & 0.3--1.2    &  0.0--1.4    &  0.3--1.3    \\
Background (statistical)           & 3.1--17.4    & 4.0--24.2   & 4.2--15.0   &  4.3--15.0   &  5.8--10.2   \\
Integrated luminosity              & 2.6          & 2.6         & 2.6         &  2.6         &  2.6         \\
\hline
Total syst.\ uncertainty (\%)   & 17.2--89.4   & 19.7--61.7  & 17.8--56.6  & 14.5--52.9   &  17.9--65.4  \\
\hline
Total stat.\ uncertainty (\%)   & 6.1--34.1    &  7.6--44.5  & 10.4--23.5  & 7.9--28.0    &  11.2--19.9  \\
\end{tabular}
}
\label{tab:syst_Zb_2bSample}
\end{table*}
\begin{table*}[htb]
\renewcommand{\arraystretch}{1.1}
\topcaption{Uncertainties (in percent) in the
differential cross sections as a function of $\DR$ and
$\Delta\phi$ between the two $\PQb$-tagged jets, $\DR$ between
the \PZ boson and the closer $\PQb$-tagged jet, and the asymmetry
$A_{\PZ\PQb\PQb}$, for the \Ztb sample.}
\centering
\begin{tabular}{lcccc}
Uncertainty (\%) & ${\rd\sigma}/{\rd\Delta\phi_{\PQb\PQb}}$ & ${\rd\sigma}/{\rd\Delta R_{{\PQb\PQb}}}$ & ${\rd\sigma}/{\rd\Delta R^{\text{min}}_{\PZ\PQb}}$ & ${\rd\sigma}/{\rd A_{{\PZ\PQb\PQb}}}$\\
\hline
JER                                & 0.8--2.0     & 1.0--5.3      & 0.6--6.1   & 0.6--4.2   \\
JES                                & 5.6--10.7    & 6.6--20.5     & 4.2--13.1  & 5.1--9.1   \\
Unfolding (MC model)               & 0.0--47.0    & 0.0--206      & 0.0--50.6  & 2.6--33.1  \\
Unfolding (MC statistics)          & 6.3--11.5    & 6.4--30.7     & 8.2--25.6  & 7.5--30.5  \\
c, udsg background model           & 0.0--3.4     & 0.0--10.3     & 0.0--14.2  & 0.0--12.3  \\
Electron (muon) efficiency         & 1.5 (2.0)    & 1.5 (2.0)     & 1.5 (2.0)  & 1.5 (2.0)  \\
$\PQb$ tagging efficiency         & 6.0          & 6.0           & 6.0        & 6.0        \\
Pileup                             & 0.4--2.4     & 1.3--3.5      & 0.5--4.6   & 1.2--6.1   \\
Background (systematic)            & 0.1--0.8     & 0.1--0.8      & 0.2--1.3   & 0.2--0.7   \\
Background (statistical)           & 3.4--5.0     & 3.7--9.4      & 3.6--15.9  & 3.3--8.8   \\
Integrated luminosity              & 2.6          & 2.6           & 2.6        & 2.6        \\
\hline
Total syst.\ uncertainty (\%)      & 13.0--50.5   &  12.5--209     & 14.2--59.5 & 13.6--47.2 \\
\hline
Total stat.\ uncertainty (\%)      & 6.9--10.1    & 7.5--17.6     & 7.4--33.1  &  6.6--18.4 \\
\end{tabular}
\label{tab:syst_Zb_2bSample_extra}
\end{table*}

\section{Results and comparison with theoretical predictions}

\subsection{Observables}
Differential cross sections as a function of a number of kinematic
observables are measured in order to characterize the production
mechanisms of \Zob events.

For \Zob events, five kinematic observables are studied. First, \pt
and $\abs{\eta}$ of the leading-\pt $\PQb$ jet are measured, together
with the \PZ boson \pt. The distributions of these variables are
directly sensitive to the $\PQb$ quark PDF and initial-state gluon
splitting and may show differences between different PDF flavour
schemes.  Searches for physics processes beyond the SM in
Lorentz-boosted topology events depend on precise knowledge of the \PZ
boson \pt distribution.  The scalar sum \HT of the transverse momenta
of all selected jets, regardless of their flavour, is related to the
structure of the hadronic system recoiling against the boson. The
measurement of this observable at high values is potentially sensitive
to the presence of intermediate heavy particles decaying hadronically,
as predicted, for example, in some SUSY scenarios.  Finally, the
topology of the system composed of the \PZ boson and $\PQb$ jet is
studied by measuring the cross section as a function of the azimuthal
angular separation between the direction of the \PZ boson and the
direction of the highest-\pt $\PQb$ jet, $\Delta\phi_{\PZ\PQb}$. This
observable is also sensitive to the presence of boosted particles
decaying into a \PZ boson and $\PQb$ quarks.

Ratios of the differential cross sections for \Zob and $\PZ$+jets
events, inclusive in the jet flavour, are also measured:
\begin{equation*}
R(x)=\frac{\rd\sigma(\PZ{+({\ge}1\PQb)})/\rd x}{\rd\sigma(\PZ{+}\text{jets})/\rd x},
\end{equation*}
with $x$ representing one of the five observables described above. The
inclusive $\PZ$+jets event selection is defined by releasing the
requirement of a b-tagged jet in the \Zob selection. In these ratios
the kinematic observables referring to the highest-\pt b-tagged jet
from the \Zob sample are used in the numerator, while for the
denominator the observables related to the highest-\pt jet from the
$\PZ$+jet sample are examined. Several systematic uncertainties cancel
in the ratios, allowing a precise comparison with theory.

For \Ztb events, the cross section is measured as a function of the
transverse momenta of the \PZ boson and of the leading and subleading
$\PQb$ jets.  In addition, the cross section is studied as a function
of several variables explicitly related to the topology of the final
state consisting of a \Z boson and the two highest-\pt $\cPqb$
jets. The invariant mass ${M_{{\PQb\PQb}}}$ of the $\PQb\PQb$ system
and the invariant mass $M_{\PZ\PQb\PQb}$ of the $\PZ\PQb\PQb$ system
are studied, because their distributions are sensitive to the presence
of heavy intermediate particles.

Angular correlations between the $\PQb$ jets and between each $\PQb$
jet and the \PZ boson are described by four observables, also studied
in Ref.~\cite{Chatrchyan:2013zjb}.  The azimuthal angular separation
$\Delta \phi_{{\PQb\PQb}}$ between the directions of the two $\PQb$
jets in the transverse plane is useful to identify back-to-back
configurations of the $\PQb$ quarks.  The distance between
the directions of the two $\PQb$ jets in the $\eta$-$\phi$ plane is
defined as $\Delta R_{{\PQb\PQb}} =
\sqrt{\smash[b]{(\Delta\eta_{{\PQb\PQb}})^{2}+(\Delta\phi_{{\PQb\PQb}})^{2}}}$,
where $\Delta \eta_{{\PQb\PQb}}$ is the separation in pseudorapidity
between the two $\PQb$ jets. This variable is sensitive to the
different production mechanisms of the \Ztb final-state
$\PQb$ quarks. In particular, it is useful to discriminate between the
contributions whose scattering amplitudes are dominated by terms
involving gluon splitting, $\cPg \to \bbbar$, and those where
a \PZ boson is emitted from one of the final-state $\PQb$ quarks. The
process $\qqbar\to \Z\bbbar$ contributes to both
cases, while $\PQq\Pg\to \Z\bbbar\mathrm{X}$ (with
$\mathrm{X}$ an additional parton) contributes to the former and
$\Pg\Pg\to \Z\bbbar$ to the latter.  These contributions
correspond, respectively, to the regions where the two $\PQb$ quarks
are almost collinear or mostly acollinear. Because two $\PQb$ jets
must be reconstructed, this measurement cannot be sensitive to
low-angle gluon splitting, where the distance between the
jet-initiating partons is smaller than twice the jet size. This region
is better explored by searching directly for pairs of b hadrons close
in space, as studied in Ref.~\cite{Chatrchyan:2013zjb}, whose decay
products might be part of a single reconstructed jet.  Another angular
observable of interest is $\Delta R_{\PZ\PQb}^{\text{min}}$, the
angular separation between the \PZ boson and the closer $\PQb$ jet in
the $\eta$-$\phi$ plane. This variable is useful for testing multileg
tree-level and NLO corrections in which a \PZ boson is radiated from a
quark, because it is sensitive to event topologies with the \PZ boson
in the vicinity of one of the two $\PQb$ jets.  Finally, the
$A_{\PZ\PQb\PQb}$ asymmetry between the $\PQb$ jet direction and the
\PZ boson direction is computed using a combination of $\Delta
R_{\PZ\PQb}^{\text{min}}$ and $\Delta
R_{\PZ\PQb}^{\text{max}}$ (the latter being the $\eta$-$\phi$
separation between the \PZ boson and the farther b
jet):
\begin{equation*}
A_{\PZ\PQb\PQb} = \frac{\Delta
R_{\PZ\PQb}^{\text{max}} - \Delta
R_{\PZ\PQb}^{\text{min}}}{\Delta
R_{\PZ\PQb}^{\text{max}} + \Delta
R_{\PZ\PQb}^{\text{min}}} .
\end{equation*}
The $A_{\PZ\PQb\PQb}$ asymmetry can provide an indirect test of pQCD
validity at higher orders of the perturbative series. A nonzero value
of $A_{\PZ\PQb\PQb}$ is related to the emission of additional gluon
radiation in the final state, while values of $A_{\PZ\PQb\PQb}$ close
to zero identify configurations in which the two $\PQb$ jets are
emitted symmetrically with respect to the \PZ boson direction.

\subsection{Theoretical predictions}
\label{sec:theory}
The measured differential cross sections for the associated
production of \PZ bosons and $\PQb$ jets are compared to several
perturbative QCD theoretical calculations.

In pQCD the amplitude for this process can be computed using two
alternative approaches. In the four-flavour scheme
(4FS)~\cite{Cordero:2009kv}, the $\PQb$ quark mass is explicitly
included in the predictions and acts as an infrared cutoff, partly
removing possible divergences in the matrix element calculation. This
approach corresponds to an effective QCD theory, with $n_{f}=4$ quark
flavours involved in the computation of the running of the strong
coupling constant \alpS. In this scheme no $\PQb$ quark PDF is used,
and the $\PQb$ quark is always produced explicitly by the gluon
splitting $\Pg \to \bbbar$ process. In the
5FS~\cite{Campbell:2003dd} (where $n_{f}=5$), the gluon splitting
contribution is included within the $\PQb$ quark PDF, and the $\PQb$
quark mass is set to zero in the matrix element calculation. The two
schemes can be defined in such a way as to provide identical
results when all orders in pQCD are computed. However, differences
appear in fixed-order predictions, where the different ordering of
terms in the matrix element expansion gives different results. The
comparison of different flavour schemes is interesting because, in
pQCD, the evolution of the $\PQb$ quark PDF as a function of the
Bjorken $x$ variable shows sizeable differences between tree-level
calculations and those at NLO. These differences are introduced by
singularities in the Altarelli--Parisi splitting functions that are
present only at NLO; they have no impact on the tree-level evolution
of the $\PQb$ quark PDF~\cite{Maltoni:2012pa}.

While NLO calculations are now available for both flavour
schemes, LO calculations are still interesting to study because they
allow the inclusion of multiple additional light, hard partons in the
matrix element. This feature is expected to provide a better
description of the real hard radiation, compared to fixed-order NLO
calculations matched with parton showering.

The \MADGRAPH plus \PYTHIA6 event generator, introduced in
Section~\ref{sec:evsim}, describes signal events at full
detector simulation level and provides theoretical predictions at
tree level for the associated production of \PZ bosons and jets,
including $\PQb$ jets. This calculation is based on the 5FS using the
LO \MADGRAPH 5.1.3.30 matrix element generator, with up to four
additional partons in the matrix element calculation. The
factorization and renormalization scales are chosen on an
event-by-event basis as the transverse mass of the event, clustered
with the \kt algorithm down to a 2$\to$2 topology, and \kt computed at
each vertex splitting,
respectively~\cite{Alwall:2008qv,Alwall:2007fs}. The matrix element
calculation is interfaced with \PYTHIA version~6.424, using tune
Z2* for parton showering, hadronization, and description of
MPI. The CTEQ6L1 PDF is adopted in the calculations.  The Drell-Yan
inclusive cross section is rescaled to the NNLO calculation provided
by \FEWZ 3.1~\cite{Gavin:2010az,Li:2012wna}, which has a uncertainty
of about 5\%. This uncertainty is not propagated into the figures
presented below.

Theoretical predictions at tree level based on \MADGRAPH matrix
elements for the $\Z + 2\PQb$ process are also computed using
the 4FS MSTW2008 LO PDF set~\cite{Martin:2009iq}.  The factorization
and renormalization scales are defined as in the 5FS case. Also in
this case, parton showering and hadronization are provided by \PYTHIAS
with the tune Z2*. The inclusive cross section is rescaled to the
$\Z + 2\PQb$ NLO calculation with
\AMCATNLO~\cite{Alwall:2014hca} for the 4FS, which has an estimated
theoretical uncertainty of 15\%, dominated by scale variations. This
uncertainty is not shown in the figures.

The event generator \AMCATNLO~\cite{Alwall:2014hca} version 2.2.1 is
used to provide results at NLO, combining matrix elements for zero,
one, and two additional partons through the {\sc FxFx}
algorithm~\cite{Frederix:2012ps}. The NNPDF 3.0 NLO PDF
set~\cite{Ball:2014uwa}, based on the 5FS, is used. Parton showering
and hadronization are performed by \PYTHIA version
8.205~\cite{Sjostrand:2007gs}, using the CUETP8M1
tune~\cite{Khachatryan:2015pea}.  The choice of QCD scales is the same
as for the LO \MADGRAPH prediction. This is the same event generator
that is used in Section~\ref{sec:evsim} to study the systematic
uncertainty in the secondary vertex mass distribution.

The 5FS is also used to compute the NLO \POWHEG prediction for a \PZ
boson associated with two extra partons, including $\PQb$
quarks~\cite{Campbell:2013vha}. Lower parton multiplicities are
described in the matrix element as well, but with no guarantee of
NLO accuracy. The scale choice is based on the \textsc{
minlo} approach~\cite{Hamilton:2012np}. The NNPDF 3.0 PDF
set~\cite{Ball:2014uwa} is used, and the matrix element calculation is
matched with the \PYTHIAE parton shower evolution and hadronization,
using the CUETP8M1 tune.

For both \AMCATNLO and \POWHEG, no further rescaling of the native
cross section is made. Theoretical systematic uncertainties in the
predictions, caused by the choice of the QCD factorization and
renormalization scales and by the propagation of the uncertainties in
PDFs, are computed. The former are estimated by varying the QCD scales
by factors of 2 and 0.5, while the latter are computed according to
PDF authors' prescriptions. The uncertainty from varying the QCD scales is
generally the dominant contribution. These theoretical uncertainties
are displayed in the figures only in the ratio plots, with the
statistical uncertainty shown separately, and add up to about 10\% and
20\% for the two calculations, respectively. For LO calculations, only
the statistical uncertainty of theoretical predictions is shown.

\subsection{Comparison with data}
The measured differential cross sections, unfolded for detector
effects, are compatible for the two leptonic channels, and are
therefore combined into an uncertainty-weighted average for a single
lepton flavour. Correlations between systematic uncertainties for the
electron and muon channels are taken into account in the
combination. In particular, all uncertainties are treated as fully
correlated, with the exception of those related to lepton
efficiencies, \ttbar background estimates, and the statistical part of
the subtraction of the c quark and udsg components from $\PZ$+jets,
and the statistical part of the unfolding uncertainty, which are
treated as fully uncorrelated. All the cross sections are measured in
the fiducial phase space defined at the generated particle level for
the unfolding procedure, as discussed in
Section~\ref{sec:unfolding}. No attempt is made to disentangle $\PQb$
jet production in the hard scattering processes and in MPI.

The integral of the unfolded distributions gives the fiducial cross section,
for a single lepton type, for the production of \Zob events,
\begin{equation*}
\sigma_{\text{fid}}( \Pp\Pp\to \PZ +({\geq}1\PQb) )
= 3.55 \pm 0.12\stat \pm 0.21\syst\unit{pb} ,
\end{equation*}
and \Ztb events,
\begin{equation*}
\sigma_{\text{fid}}( \Pp\Pp\to \PZ +({\geq}2\PQb))
= 0.331 \pm 0.011\stat \pm 0.035\syst\unit{pb} .
\end{equation*}
These measured values can be compared with the corresponding
predictions at NLO of \AMCATNLO interfaced with \PYTHIAE (described in
Sec.\ref{sec:theory}), $4.23^{+0.27}_{-0.37}$\unit{pb} for \Zob and
$0.356^{+0.030}_{-0.031}$\unit{pb} for \Ztb. The prediction
overestimates by about 20\% the measured value for \Zob, while a
reasonable agreement is found for \Ztb within uncertainties.

The ratio of the cross sections in the fiducial phase space for the
production of at least two and at least one $\PQb$ jet is
\begin{equation*}
\frac{\sigma_{\text{fid}}( \Pp\Pp\to \PZ +({\geq}2\PQb))}{\sigma_{\text{fid}}( \Pp\Pp\to \PZ
+({\geq}1\PQb))} = 0.093 \pm 0.004\stat \pm 0.007\syst ,
\end{equation*}
to be compared with the theoretical prediction
$0.084^{+0.002}_{-0.001}$ where
the systematic uncertainties are considered as fully correlated.

\begin{figure*}[hbt]
\centering
\includegraphics[width=\ghmFigWidth]{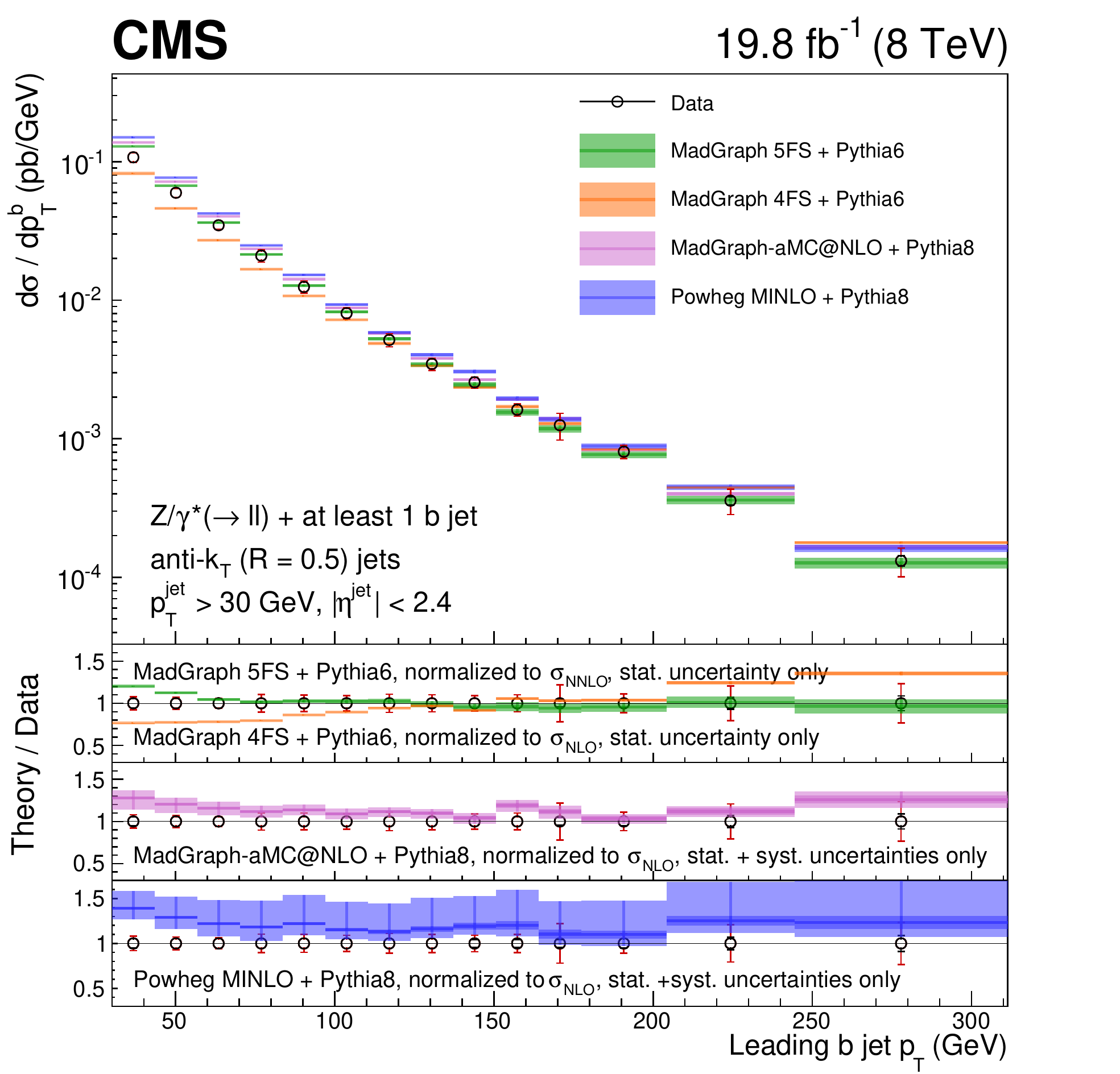}
\includegraphics[width=\ghmFigWidth]{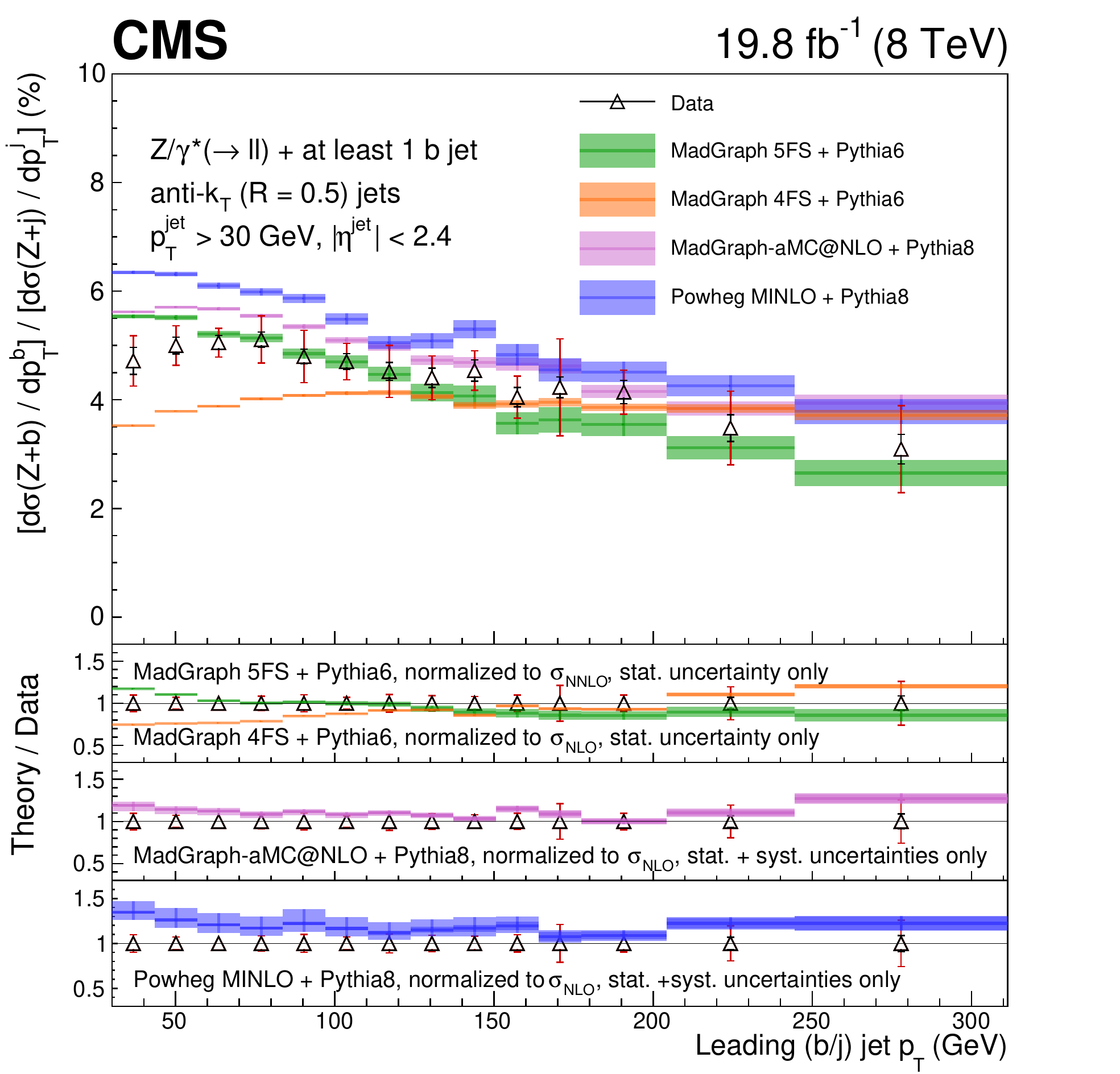}
\caption{Differential fiducial cross section for Z(1b) production as a
  function of the leading $\PQb$ jet \pt (left), and the cross section
  ratio for Z(1b) and Z+jets production as a function of the leading
  $\PQb$/inclusive (j) jet \pt (right), compared with the \MADGRAPH
  5FS, \MADGRAPH 4FS, \AMCATNLO, and \POWHEG\ \textsc{minlo} theoretical
  predictions (shaded bands), normalized to the theoretical cross
  sections described in the text.  For each data point the statistical
  and the total (sum in quadrature of statistical and systematic)
  uncertainties are represented by the double error bar.  The width of
  the shaded bands represents the uncertainty in the theoretical
  predictions, and, for NLO calculations, the inner darker area
  represents the statistical component only.
\label{fig:w_first_bjet_pt_unfolding}}
\end{figure*}

\begin{figure*}[hbt]
\centering
\includegraphics[width=\ghmFigWidth]{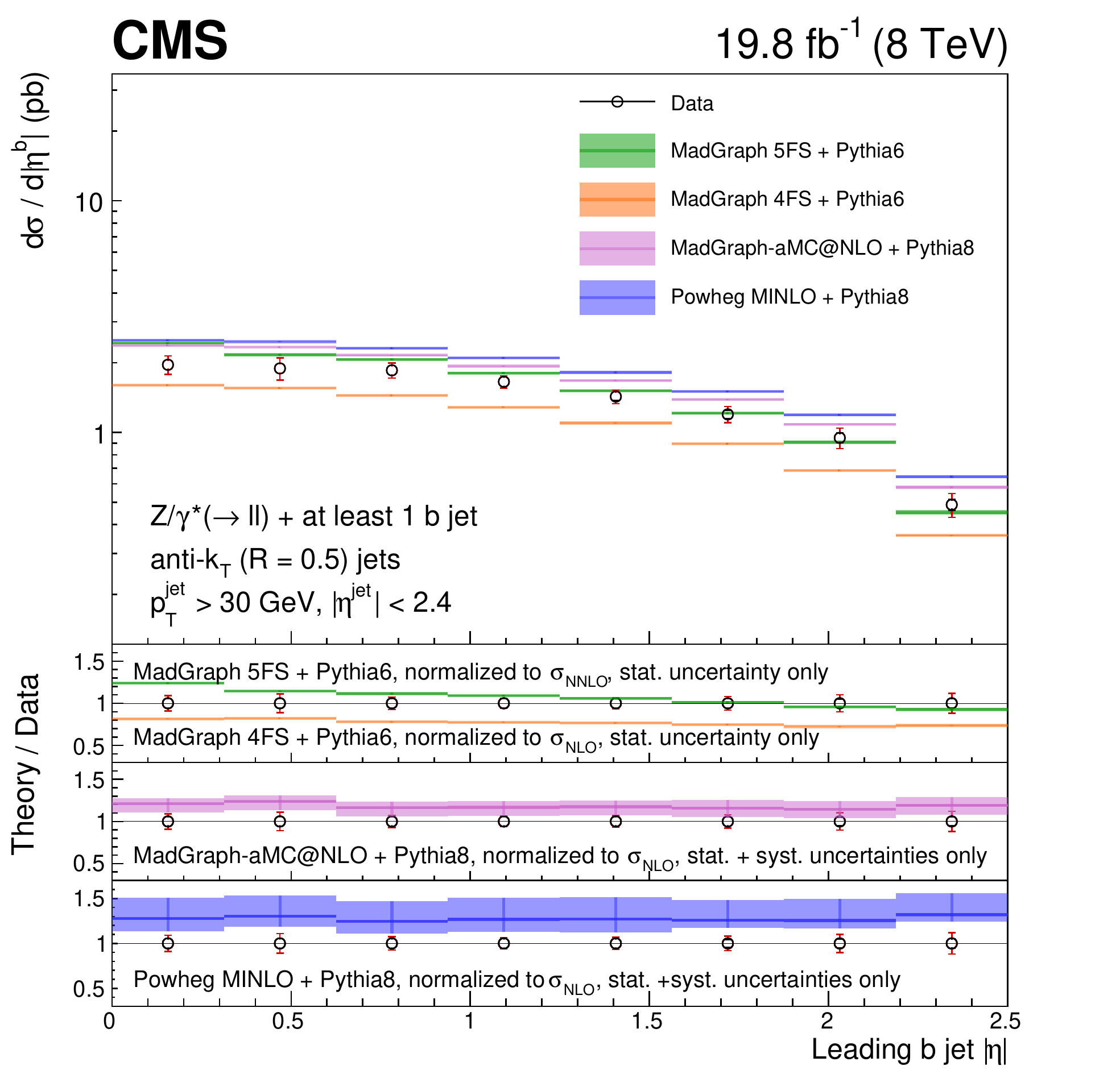}
\includegraphics[width=\ghmFigWidth]{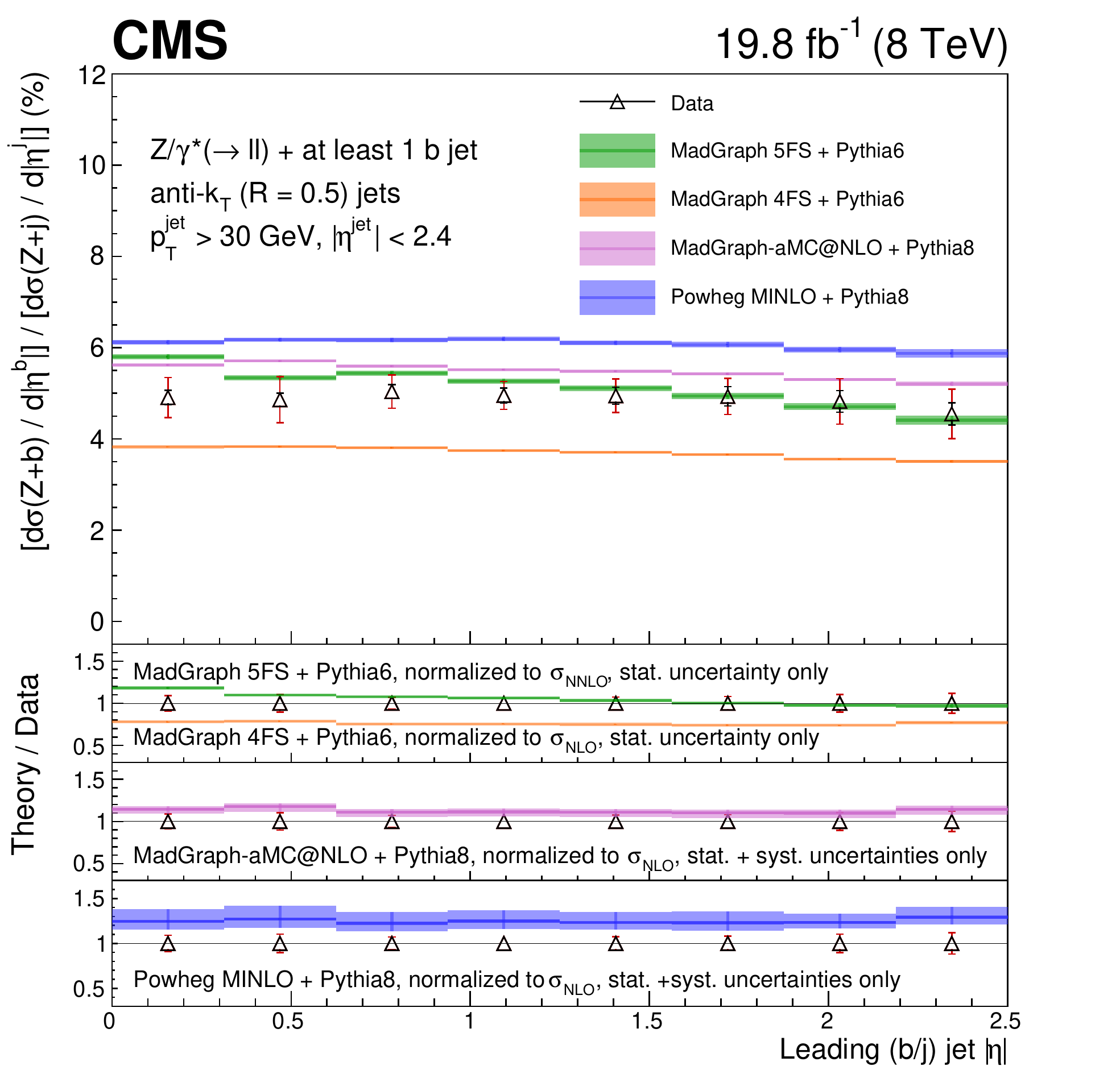}
\caption{Differential fiducial cross section for Z(1b) production as a
  function of the leading $\PQb$ jet $\abs{\eta}$ (left), and the
  cross section ratio for Z(1b) and Z+jets production as a function of
  the leading $\PQb$/inclusive (j) jet $\abs{\eta}$ (right), compared
  with the \MADGRAPH 5FS, \MADGRAPH 4FS, \AMCATNLO, and \POWHEG\ \textsc{minlo} theoretical predictions (shaded bands), normalized to the
  theoretical cross sections described in the text.  For each data
  point the statistical and the total (sum in quadrature of
  statistical and systematic) uncertainties are represented by the
  double error bar.  The width of the shaded bands represents the
  uncertainty in the theoretical predictions, and, for NLO
  calculations, theoretical systematic uncertainties are added in the
  ratio plots with the inner darker area representing the statistical
  component only.
\label{fig:w_first_bjet_eta_abs_unfolding}}
\end{figure*}

\begin{figure*}[hbt]
\centering
\includegraphics[width=\ghmFigWidth]{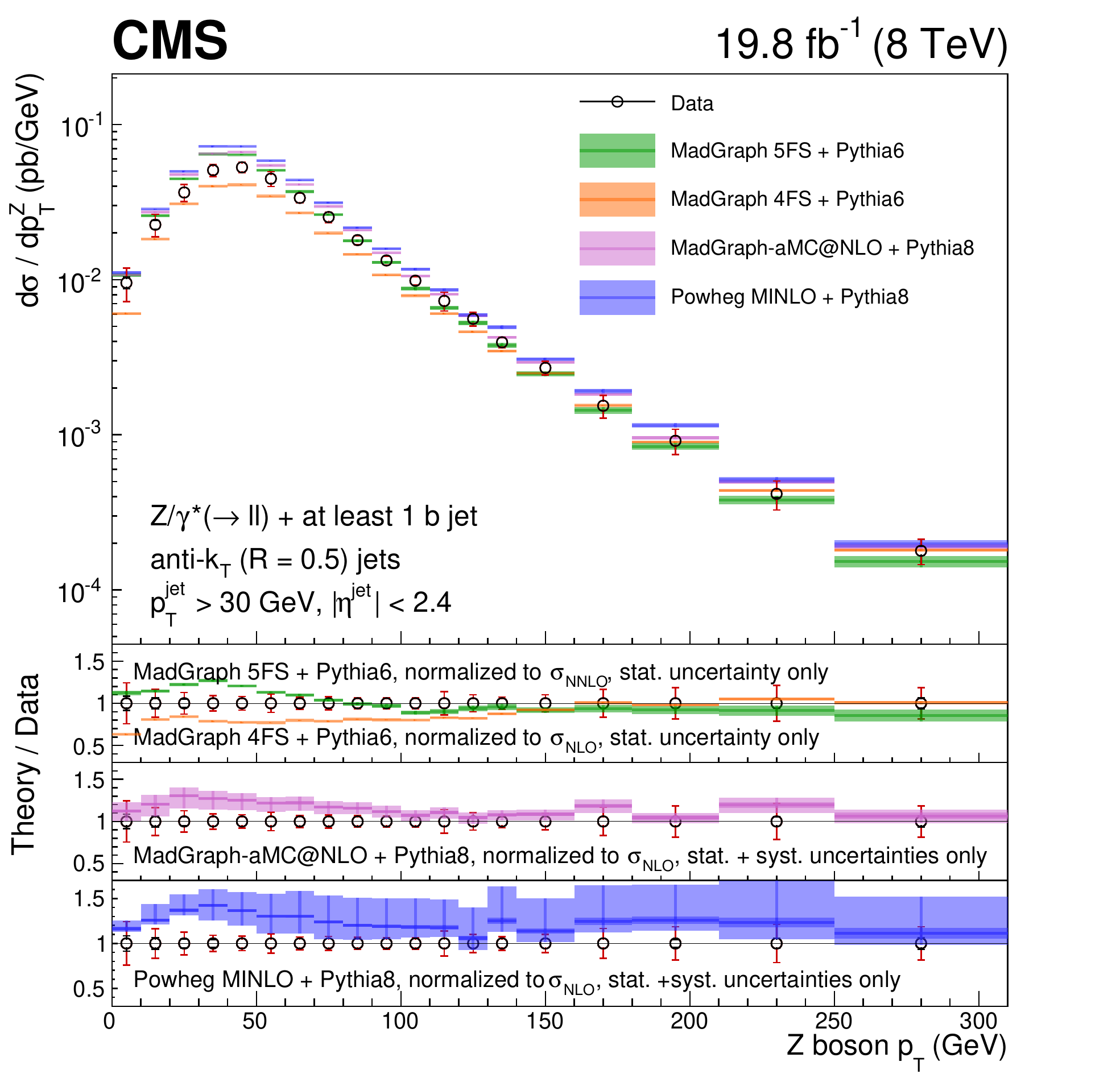}
\includegraphics[width=\ghmFigWidth]{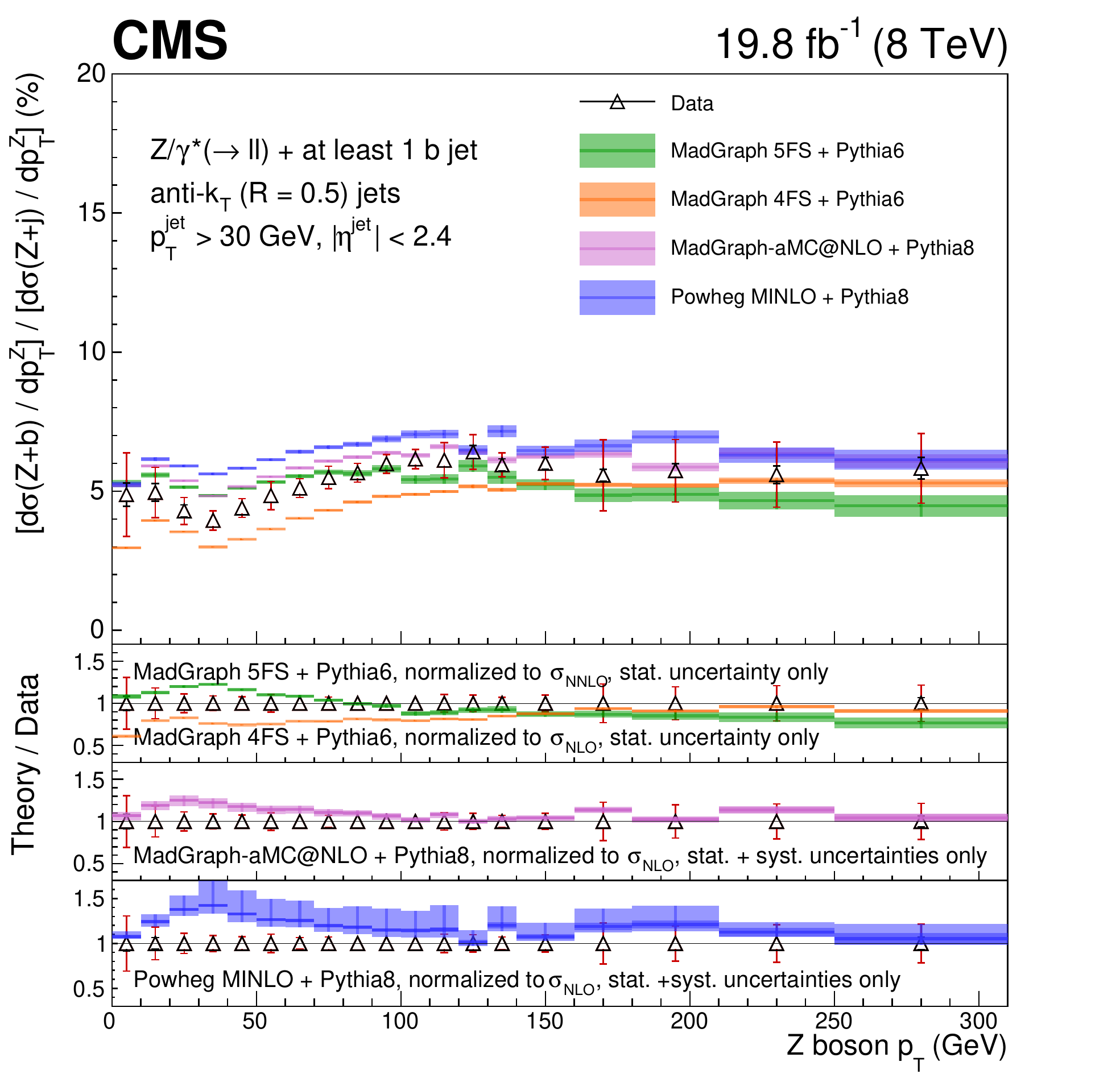}
\caption{Differential fiducial cross section for Z(1b) production as a
  function of the \PZ boson \pt (left), and the cross section ratio
  for Z(1b) and Z+jets production as a function of the \PZ boson \pt
  (right), compared with the \MADGRAPH 5FS, \MADGRAPH 4FS, \AMCATNLO,
  and \POWHEG\ \textsc{minlo} theoretical predictions (shaded bands),
  normalized to the theoretical cross sections described in the text.
  For each data point the statistical and the total (sum in quadrature
  of statistical and systematic) uncertainties are represented by the
  double error bar.  The width of the shaded bands represents the
  uncertainty in the theoretical predictions, and, for NLO
  calculations, theoretical systematic uncertainties are added in the
  ratio plots with the inner darker area representing the statistical
  component only.
\label{fig:w_pt_Z_b_unfolding}}
\end{figure*}

\begin{figure*}[hbt]
\centering
\includegraphics[width=\ghmFigWidth]{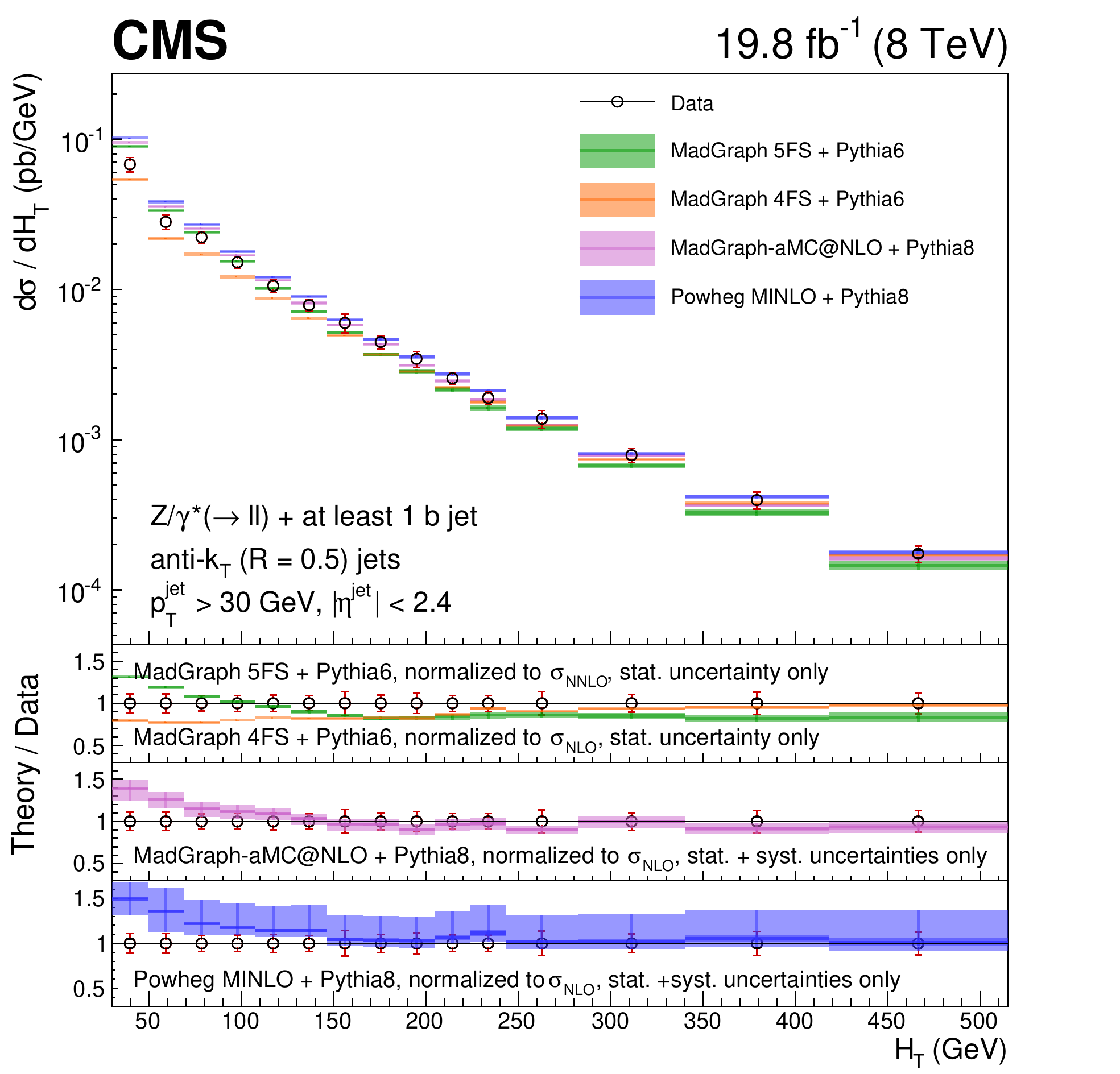}
\includegraphics[width=\ghmFigWidth]{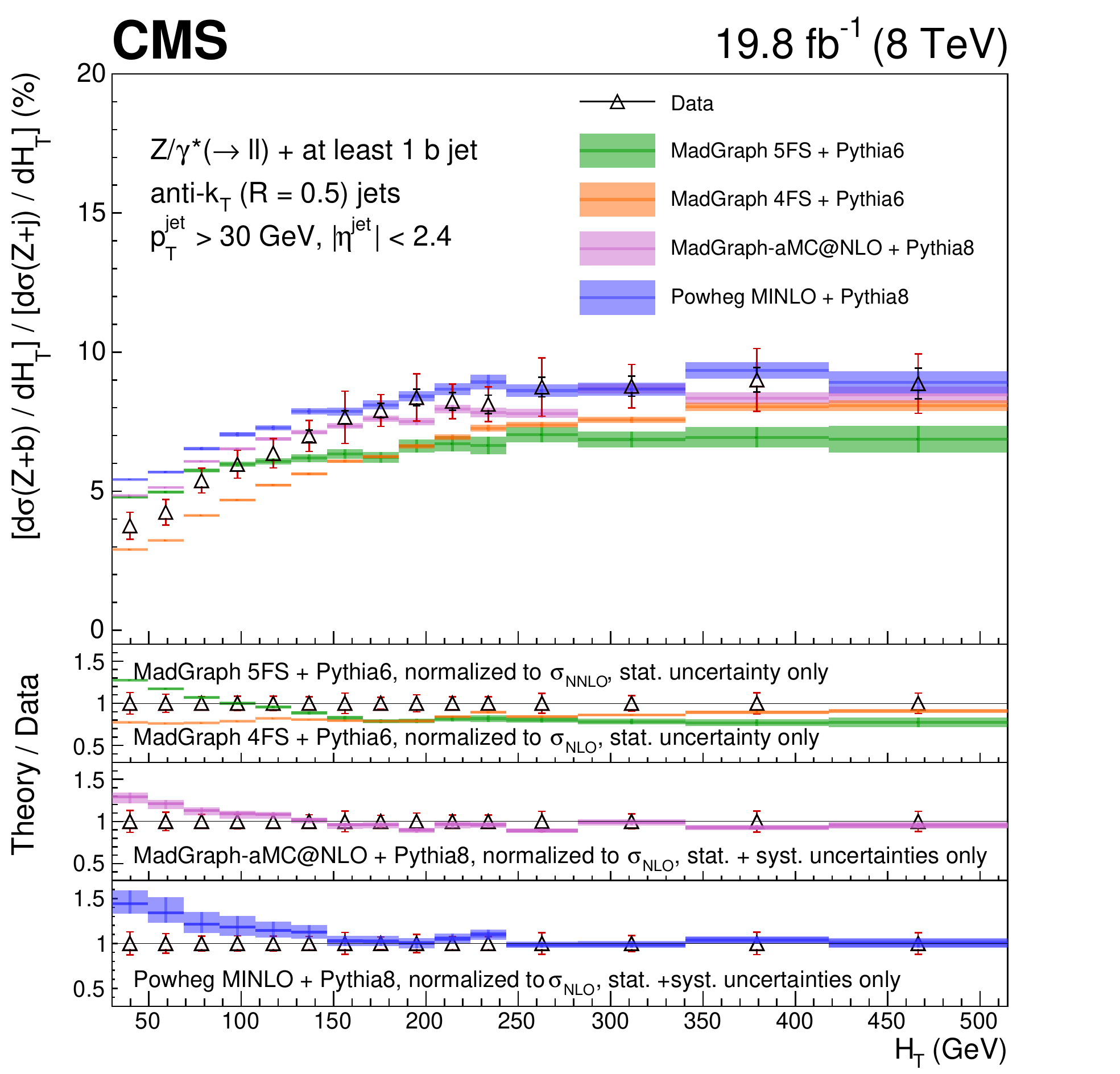}
\caption{Differential fiducial cross section for Z(1b) production as a
  function of \HT (left), and the cross section ratio for Z(1b) and
  Z+jets production as a function of \HT (right), compared with the
  \MADGRAPH 5FS, \MADGRAPH 4FS, \AMCATNLO, and \POWHEG\ \textsc{minlo}
  theoretical predictions (shaded bands), normalized to the
  theoretical cross sections described in the text.  For each data
  point the statistical and the total (sum in quadrature of
  statistical and systematic) uncertainties are represented by the
  double error bar.  The width of the shaded bands represents the
  uncertainty in the theoretical predictions, and, for NLO
  calculations, theoretical systematic uncertainties are added in the
  ratio plots with the inner darker area representing the statistical
  component only.
\label{fig:w_Ht_b_unfolding}}
\end{figure*}

\begin{figure*}[hbtp]
\centering
\includegraphics[width=\ghmFigWidth]{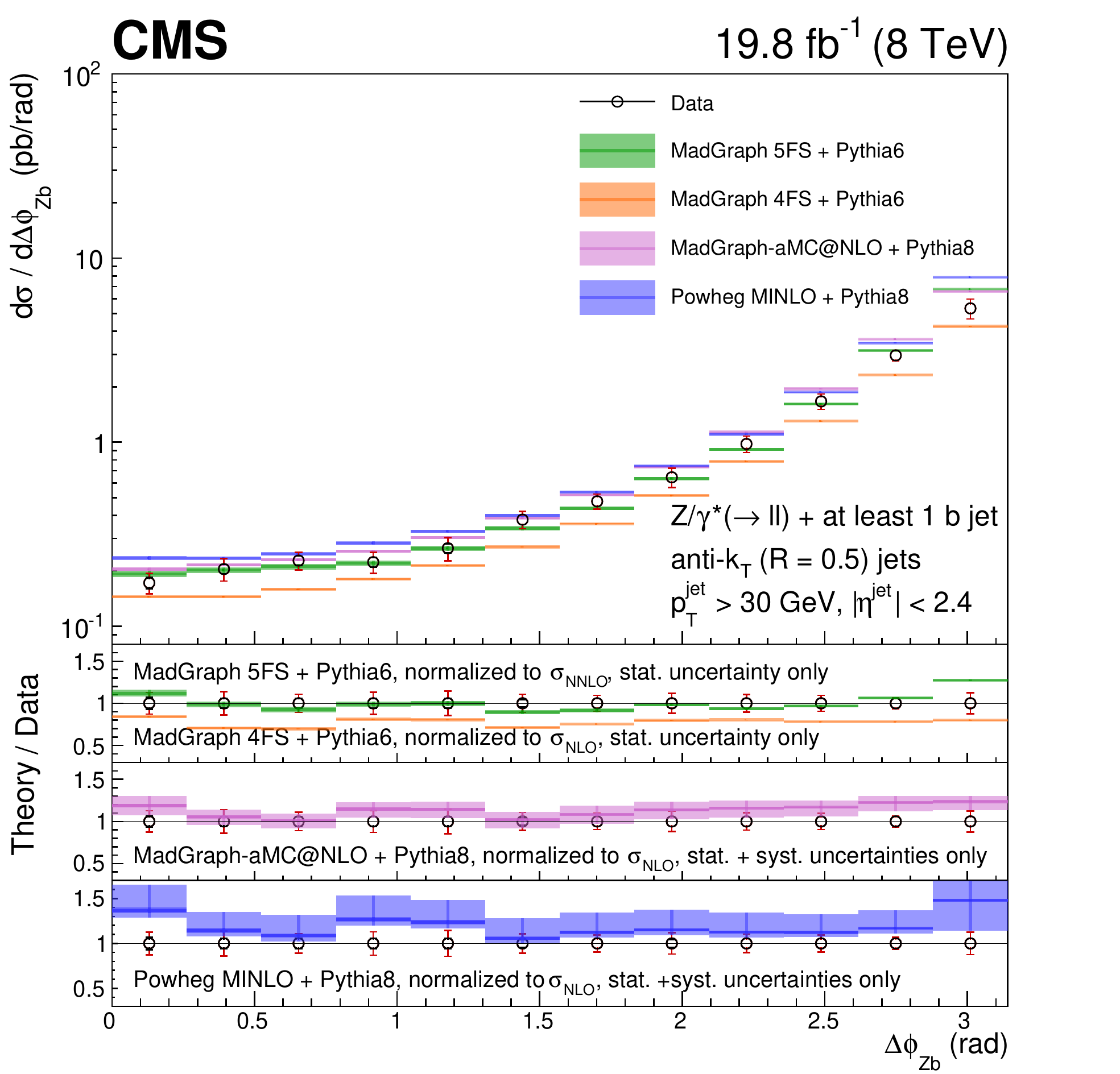}
\includegraphics[width=\ghmFigWidth]{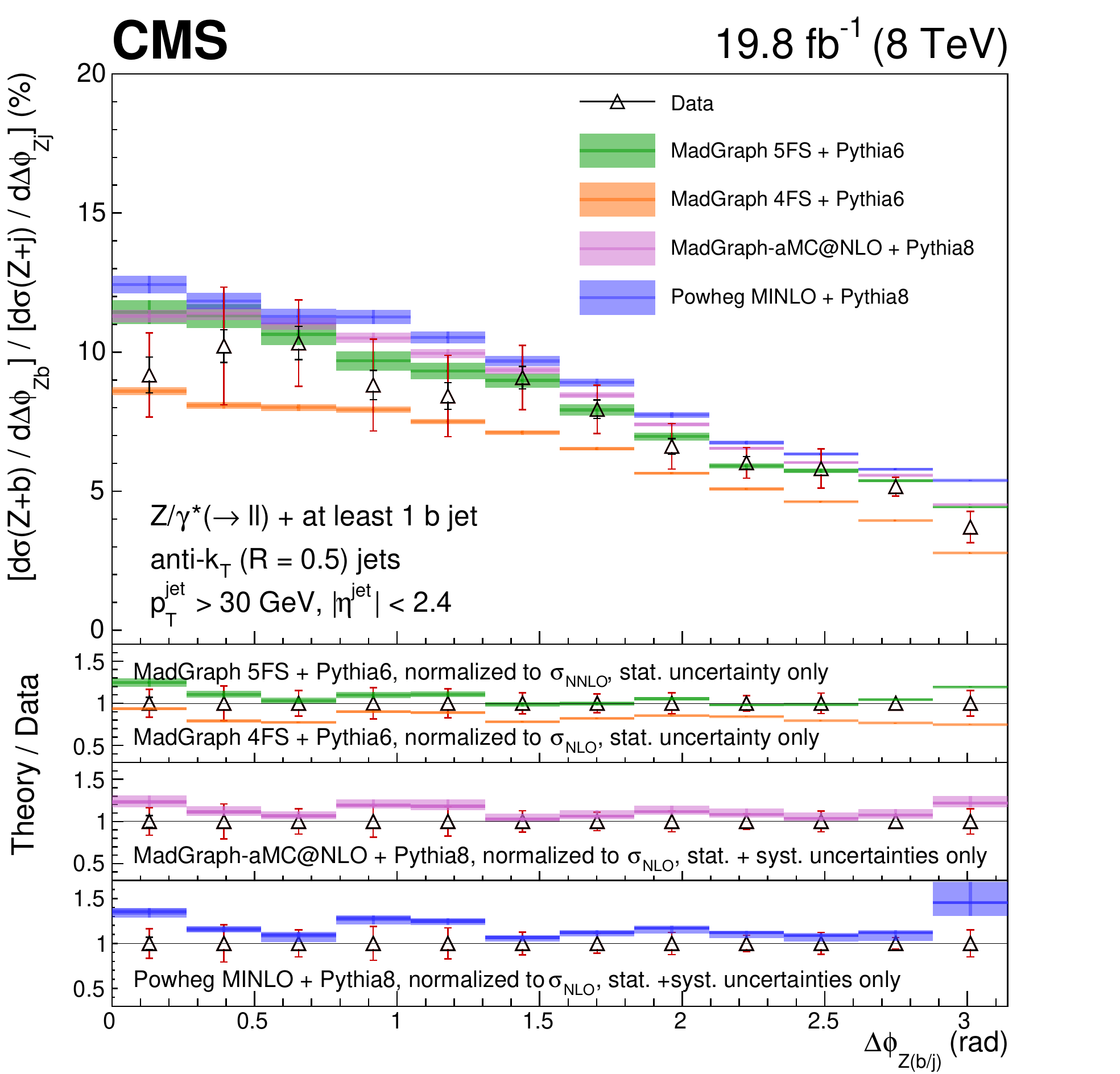}
\caption{Differential fiducial cross section for Z(1b) production as a
  function of $\Delta\phi_{\PZ\PQb}$ (left), and the cross section
  ratio for Z(1b) and Z+jets production as a function of
  $\Delta\phi_{\mathrm{Z(b/j)}}$ (right), compared with the \MADGRAPH
  5FS, \MADGRAPH 4FS, \AMCATNLO, and \POWHEG\ \textsc{minlo} theoretical
  predictions (shaded bands), normalized to the theoretical cross
  sections described in the text.  For each data point the statistical
  and the total (sum in quadrature of statistical and systematic)
  uncertainties are represented by the double error bar.  The width of
  the shaded bands represents the uncertainty in the theoretical
  predictions, and, for NLO calculations, theoretical systematic
  uncertainties are added in the ratio plots with the inner darker
  area representing the statistical component only.
\label{fig:w_delta_phi_b_unfolding}}
\end{figure*}

Results for the differential cross sections for the \Zob events are
presented in
Figs.~\ref{fig:w_first_bjet_pt_unfolding}--\ref{fig:w_delta_phi_b_unfolding},
together with the ratios to the corresponding observables for the
inclusive $\PZ$+jets event selection. Where applicable, the last bin
also includes overflow values.  A discrepancy of about 20\% is seen in
the overall normalization for the 4FS-based prediction, of the same
order of magnitude as its estimated theoretical uncertainty. The
\POWHEG prediction tends to overestimate the cross sections by about
25\%.

Apart for the normalization difference, the pQCD calculation with
massive $\PQb$ quarks (4FS) seems to reproduce, slightly better, the shape of
the observed distributions in the soft momentum regime of $\PQb$
jets. For the leading \PQb jet \pt spectrum
(Fig.~\ref{fig:w_first_bjet_pt_unfolding}), the ratio with data is
reasonably flat below 80\GeV, whereas it presents a clear slope in the
higher \pt range. A similar behaviour is clear in the \PZ boson \pt
distribution below 130\GeV (Fig.~\ref{fig:w_pt_Z_b_unfolding}) and in
the \HT spectrum below 200\GeV (Fig.~\ref{fig:w_Ht_b_unfolding}). The
\POWHEG generator considerably overestimates the soft parts of the \pt
and \HT spectra. The leading $\PQb$ jet $\abs{\eta}$ spectrum shape is
well reproduced by the \MADGRAPH 4FS configuration
(Fig.~\ref{fig:w_first_bjet_eta_abs_unfolding}), while \MADGRAPH 5FS
calculation slightly overestimates the central part of the spectrum.
The shape of the distribution of the azimuthal angular separation
$\Delta\phi_{\PZ\PQb}$ between the \PZ boson and the leading
$\PQb$ jet is reproduced within uncertainties by all the calculations
(Fig.~\ref{fig:w_delta_phi_b_unfolding}). The NLO \AMCATNLO
predictions have similar behaviour to those from LO \MADGRAPH 5FS.  As far
as the NLO \POWHEG-based prediction is concerned, it shows a similar
behaviour to \AMCATNLO, but the discrepancies are larger, reaching
about 40\% at the peak of the \PZ boson \pt spectrum.
In general, the shape predicted by each calculation compares with data,
within uncertainties, in a similar way in the high \PZ boson \pt and \HT
regions, which are potentially sensitive to new physics contributions.

The underestimation of the normalization by \MADGRAPH 4FS and the
overestimation by \POWHEG are also observed in the ratio of \Zob
and inclusive $\PZ$+jets processes (described by the \MADGRAPH
generator in the 5FS).  The pseudorapidity distribution
(Fig.~\ref{fig:w_first_bjet_eta_abs_unfolding}), with an almost flat
shape, clearly shows that the ratio for the 4FS-based prediction is
about 4\%, compared to the 5\% of the 5FS-based calculations, while
\POWHEG predicts about 6\%. The 4FS prediction also fails to reproduce
the ratio of the leading jet \pt spectra
(Fig.~\ref{fig:w_first_bjet_pt_unfolding}), which is clearly
underestimated below 80\GeV. In contrast, \POWHEG overestimates the
spectrum in the soft region by about 30\%. Similar discrepancies,
although less pronounced, are observed for \HT and the \PZ boson \pt.
The ratio as a function of the azimuthal separation between the \PZ
boson and the $\PQb$ jet (Fig.~\ref{fig:w_delta_phi_b_unfolding}) is
also slightly underestimated by the \MADGRAPH 4FS prediction when the
\PZ boson is approximately back-to-back with respect to the leading
$\PQb$ jet, with a difference in the azimuthal angles close to $\pi$.

\begin{figure}[hbt]
\centering
\includegraphics[width=\ghmFigWidthTwo]{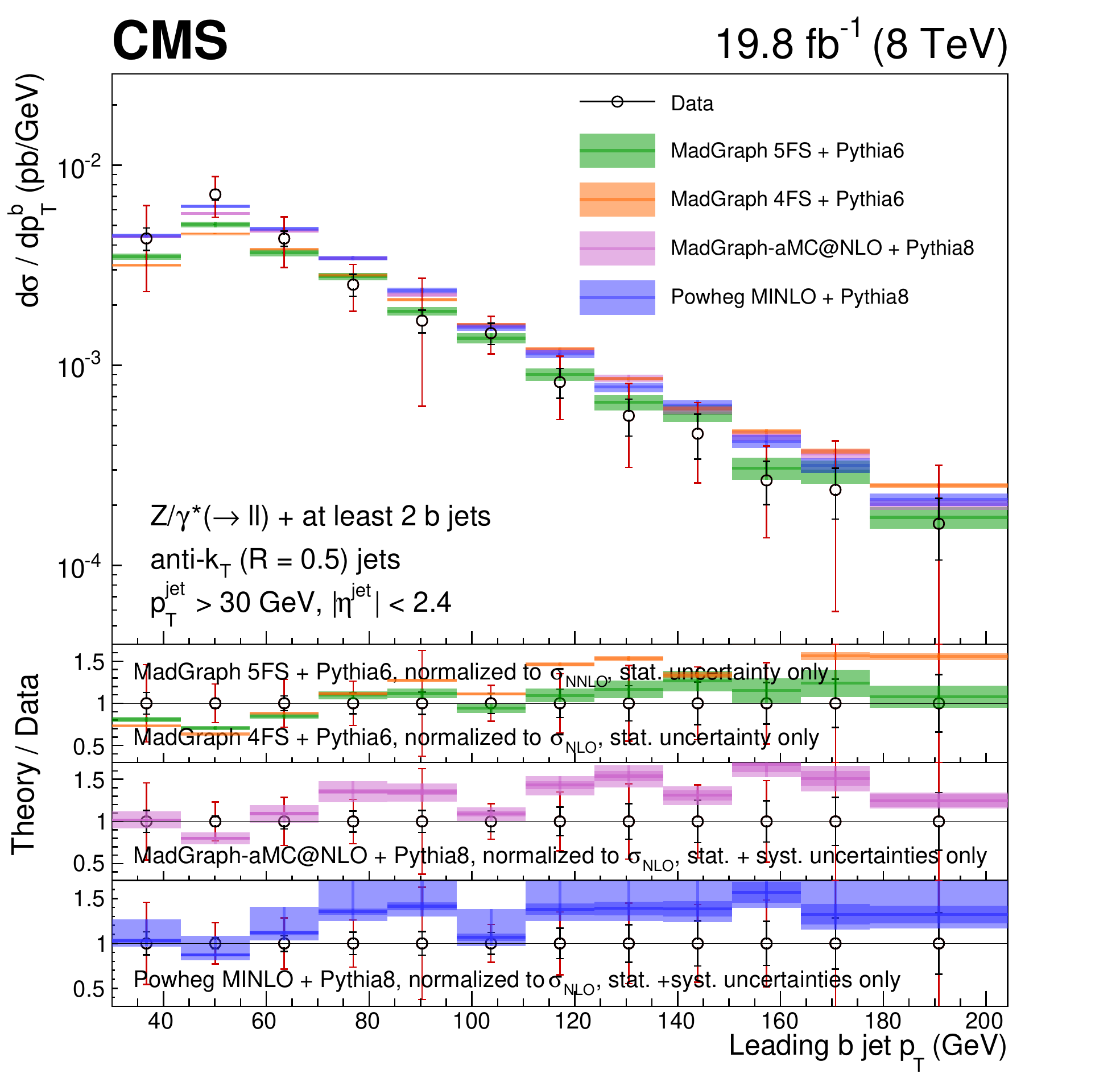}
\caption{Differential fiducial cross section for Z(2b) production as a
  function of the leading $\PQb$ jet \pt, compared with the \MADGRAPH
  5FS, \MADGRAPH 4FS, \AMCATNLO, and \POWHEG\ \textsc{minlo} theoretical
  predictions (shaded bands), normalized to the theoretical cross
  sections described in the text.  For each data point the statistical
  and the total (sum in quadrature of statistical and systematic)
  uncertainties are represented by the double error bar.  The width of
  the shaded bands represents the uncertainty in the theoretical
  predictions, and, for NLO calculations, theoretical systematic
  uncertainties are added in the ratio plots with the inner darker
  area representing the statistical component only.
\label{fig:w_first_bjet_pt_2b_unfolding}}
\end{figure}

\begin{figure}[hbt]
\centering
\includegraphics[width=\ghmFigWidthTwo]{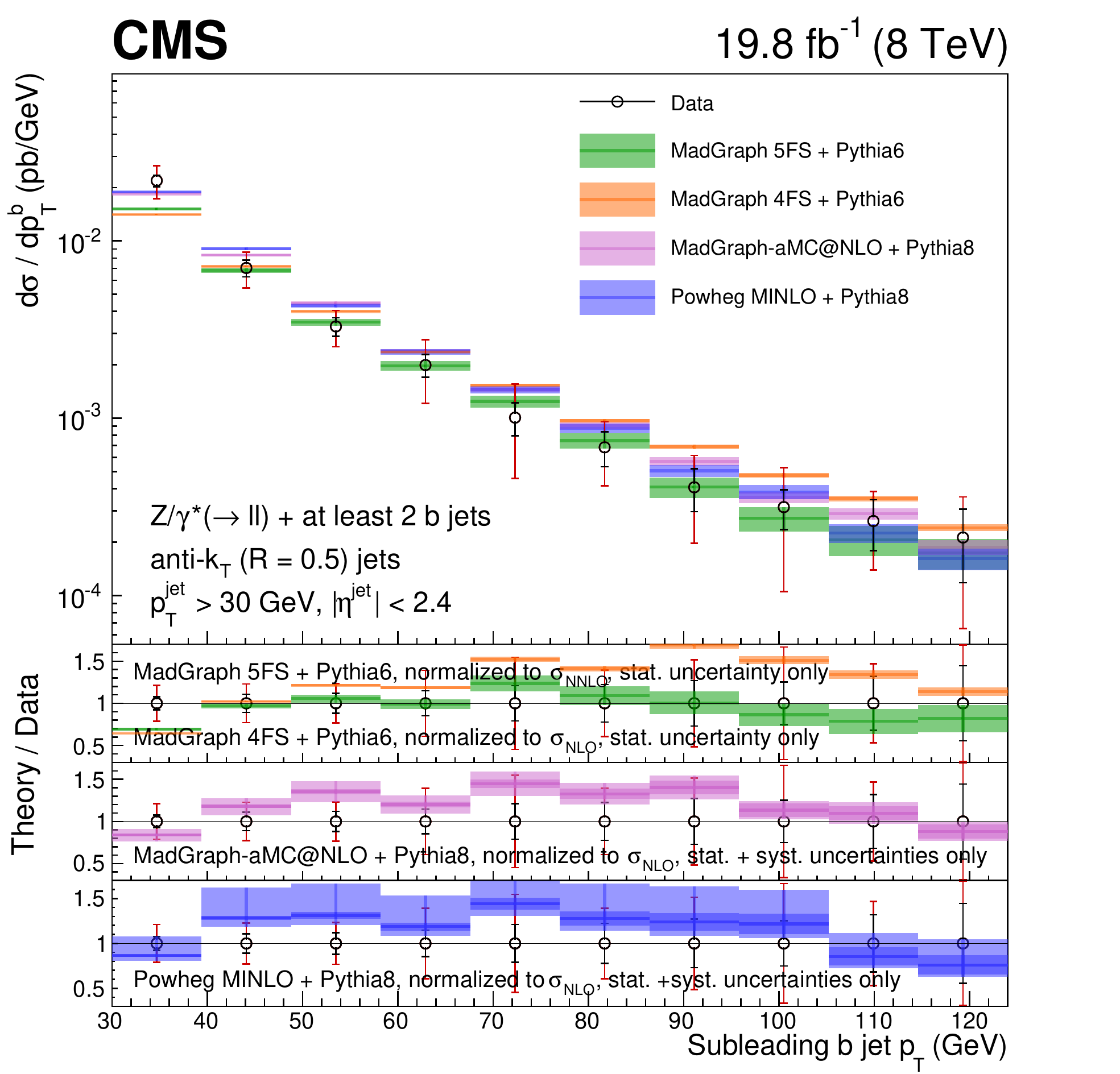}
\caption{Differential fiducial cross section for Z(2b) production as a
  function of the subleading $\PQb$ jet \pt, compared with the
  \MADGRAPH 5FS, \MADGRAPH 4FS, \AMCATNLO, and \POWHEG\ \textsc{minlo}
  theoretical predictions (shaded bands), normalized to the
  theoretical cross sections described in the text.  For each data
  point the statistical and the total (sum in quadrature of
  statistical and systematic) uncertainties are represented by the
  double error bar.  The width of the shaded bands represents the
  uncertainty in the theoretical predictions, and, for NLO
  calculations, theoretical systematic uncertainties are added in the
  ratio plots with the inner darker area representing the statistical
  component only.
\label{fig:w_second_bjet_pt_2b_unfolding}}
\end{figure}

\begin{figure}[hbt]
\centering
\includegraphics[width=\ghmFigWidthTwo]{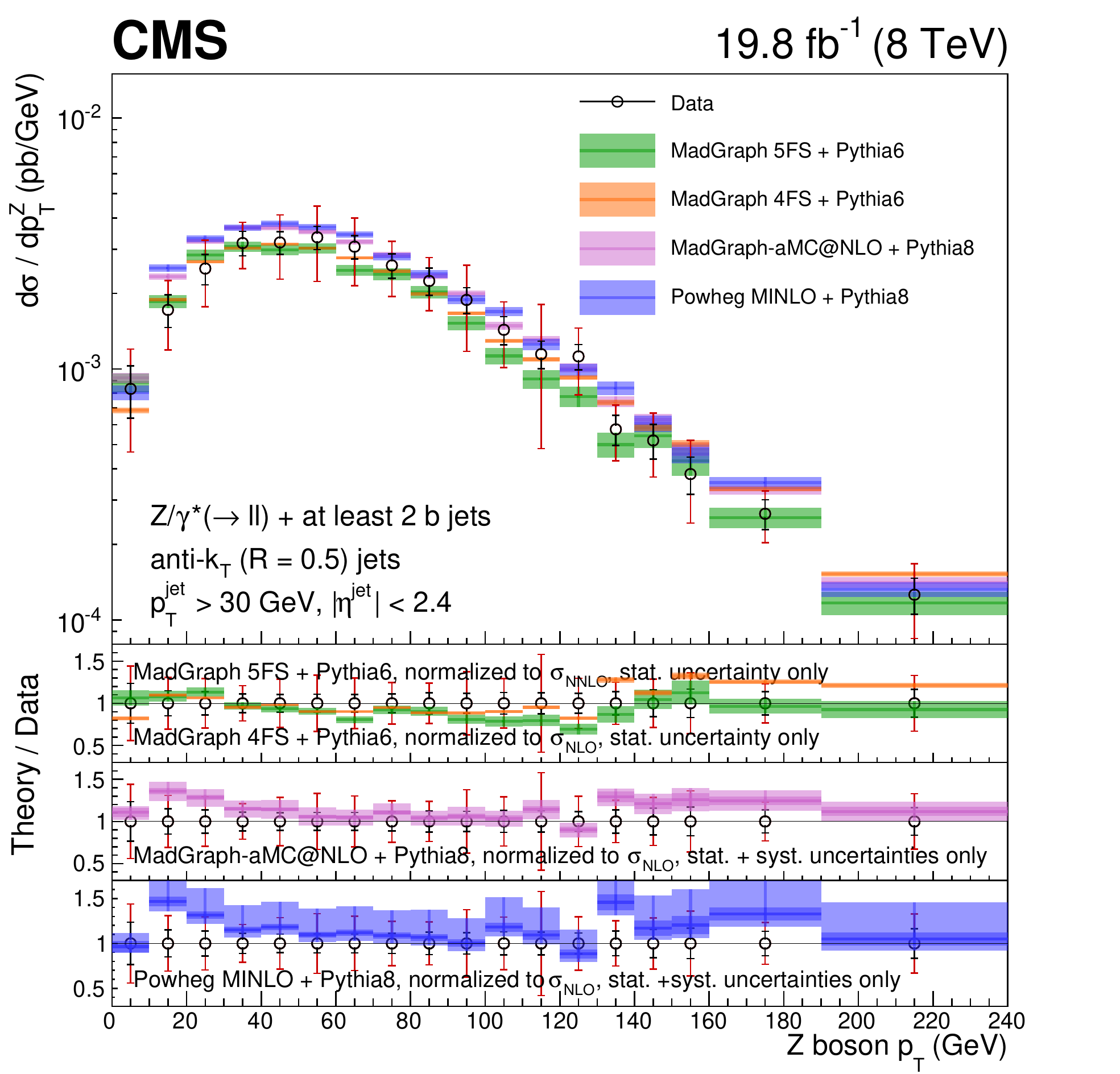}
\caption{Differential fiducial cross section for Z(2b) production as a
  function of the \PZ boson \pt, compared with the \MADGRAPH 5FS,
  \MADGRAPH 4FS, \AMCATNLO, and \POWHEG\ \textsc{minlo} theoretical
  predictions (shaded bands), normalized to the theoretical cross
  sections described in the text.  For each data point the statistical
  and the total (sum in quadrature of statistical and systematic)
  uncertainties are represented by the double error bar.  The width of
  the shaded bands represents the uncertainty in the theoretical
  predictions, and, for NLO calculations, theoretical systematic
  uncertainties are added in the ratio plots with the inner darker
  area representing the statistical component only.
\label{fig:w_pt_Z_b_2b_unfolding}}
\end{figure}

\begin{figure}[hbt]
\centering
\includegraphics[width=\ghmFigWidthTwo]{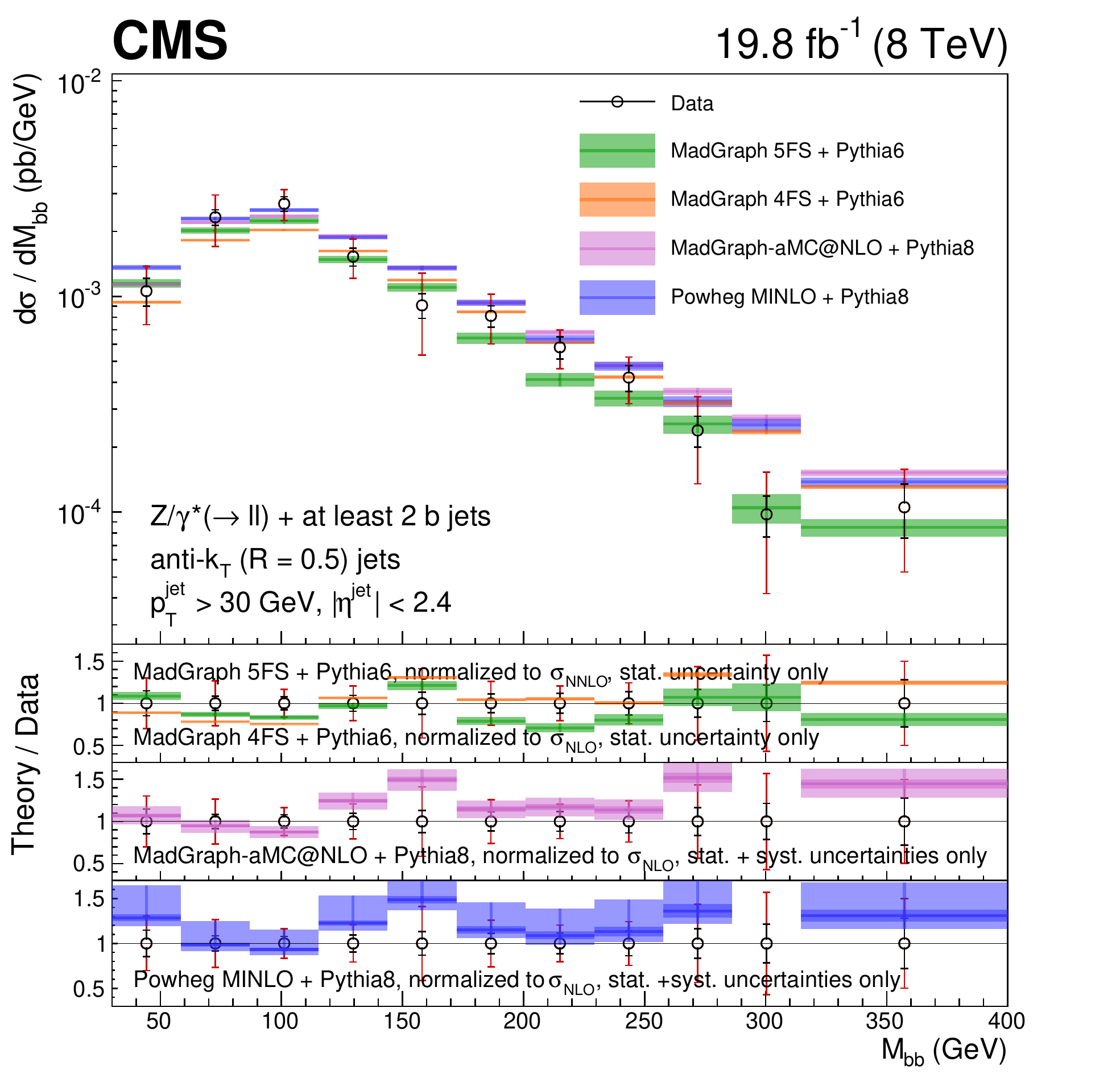}
\caption{Differential fiducial cross section for Z(2b) production as a
  function of the invariant mass of the $\PQb$ jet pair,
  $M_{\PQb\PQb}$, compared with the \MADGRAPH 5FS, \MADGRAPH 4FS,
  \AMCATNLO, and \POWHEG\ \textsc{minlo} theoretical predictions (shaded
  bands), normalized to the theoretical cross sections described in
  the text.  For each data point the statistical and the total (sum in
  quadrature of statistical and systematic) uncertainties are
  represented by the double error bar.  The width of the shaded bands
  represents the uncertainty in the theoretical predictions, and, for
  NLO calculations, theoretical systematic uncertainties are added in
  the ratio plots with the inner darker area representing the
  statistical component only.
\label{fig:w_bb_mass_unfolding}}
\end{figure}

\begin{figure}[hbt]
\centering
\includegraphics[width=\ghmFigWidthTwo]{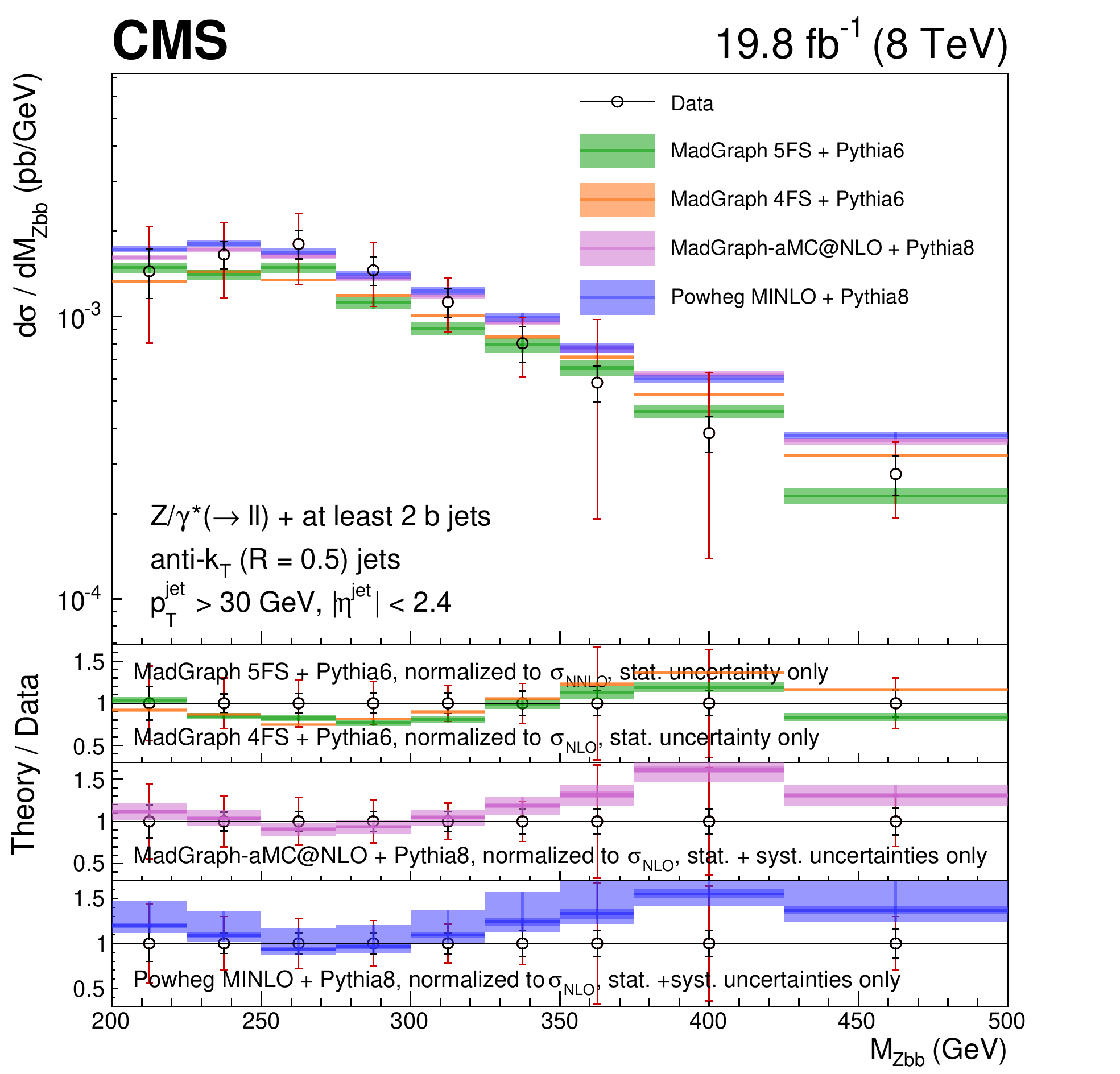}
\caption{Differential fiducial cross section for Z(2b) production as a
  function of the invariant mass of the $\PZ\PQb\PQb$ system,
  $M_\PZ\PQb\PQb$, compared with the \MADGRAPH 5FS, \MADGRAPH 4FS,
  \AMCATNLO, and \POWHEG\ \textsc{minlo} theoretical predictions (shaded
  bands), normalized to the theoretical cross sections described in
  the text.  For each data point the statistical and the total (sum in
  quadrature of statistical and systematic) uncertainties are
  represented by the double error bar.  The width of the shaded bands
  represents the uncertainty in the theoretical predictions, and, for
  NLO calculations, theoretical systematic uncertainties are added in
  the ratio plots with the inner darker area representing the
  statistical component only.
\label{fig:w_Zbb_mass_unfolding}}
\end{figure}

\begin{figure}[hbt]
\centering
\includegraphics[width=\ghmFigWidthTwo]{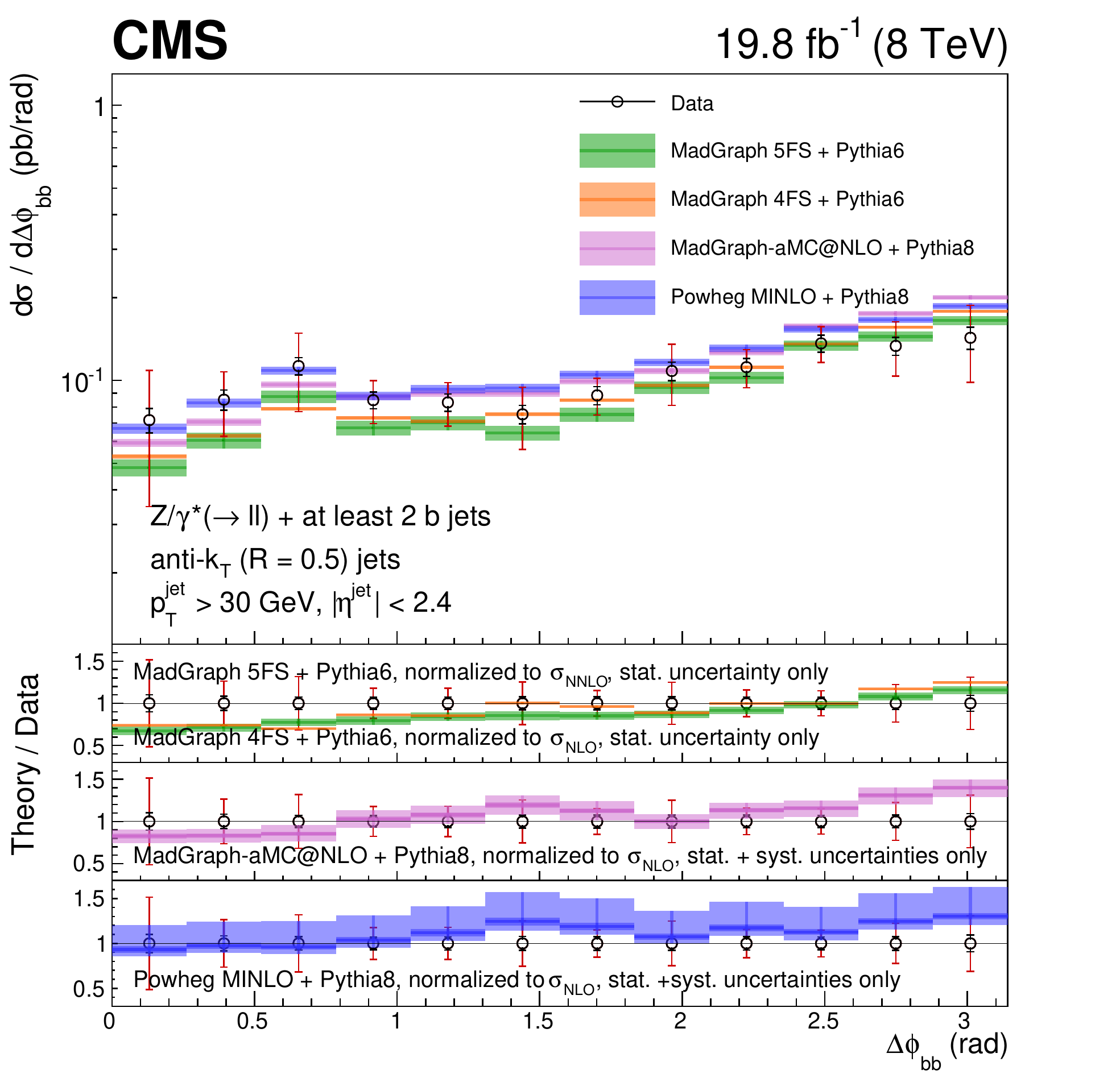}
\caption{Differential fiducial cross section for Z(2b) production as a
  function of $\Delta \phi_{\PQb\PQb}$, compared with the \MADGRAPH
  5FS, \MADGRAPH 4FS, \AMCATNLO, and \POWHEG\ \textsc{minlo} theoretical
  predictions (shaded bands), normalized to the theoretical cross
  sections described in the text.  For each data point the statistical
  and the total (sum in quadrature of statistical and systematic)
  uncertainties are represented by the double error bar.  The width of
  the shaded bands represents the uncertainty in the theoretical
  predictions, and, for NLO calculations, theoretical systematic
  uncertainties are added in the ratio plots with the inner darker
  area representing the statistical component only.
\label{fig:w_delta_phi_2b_unfolding}}
\end{figure}

\begin{figure}[hbt]
\centering
\includegraphics[width=\ghmFigWidthTwo]{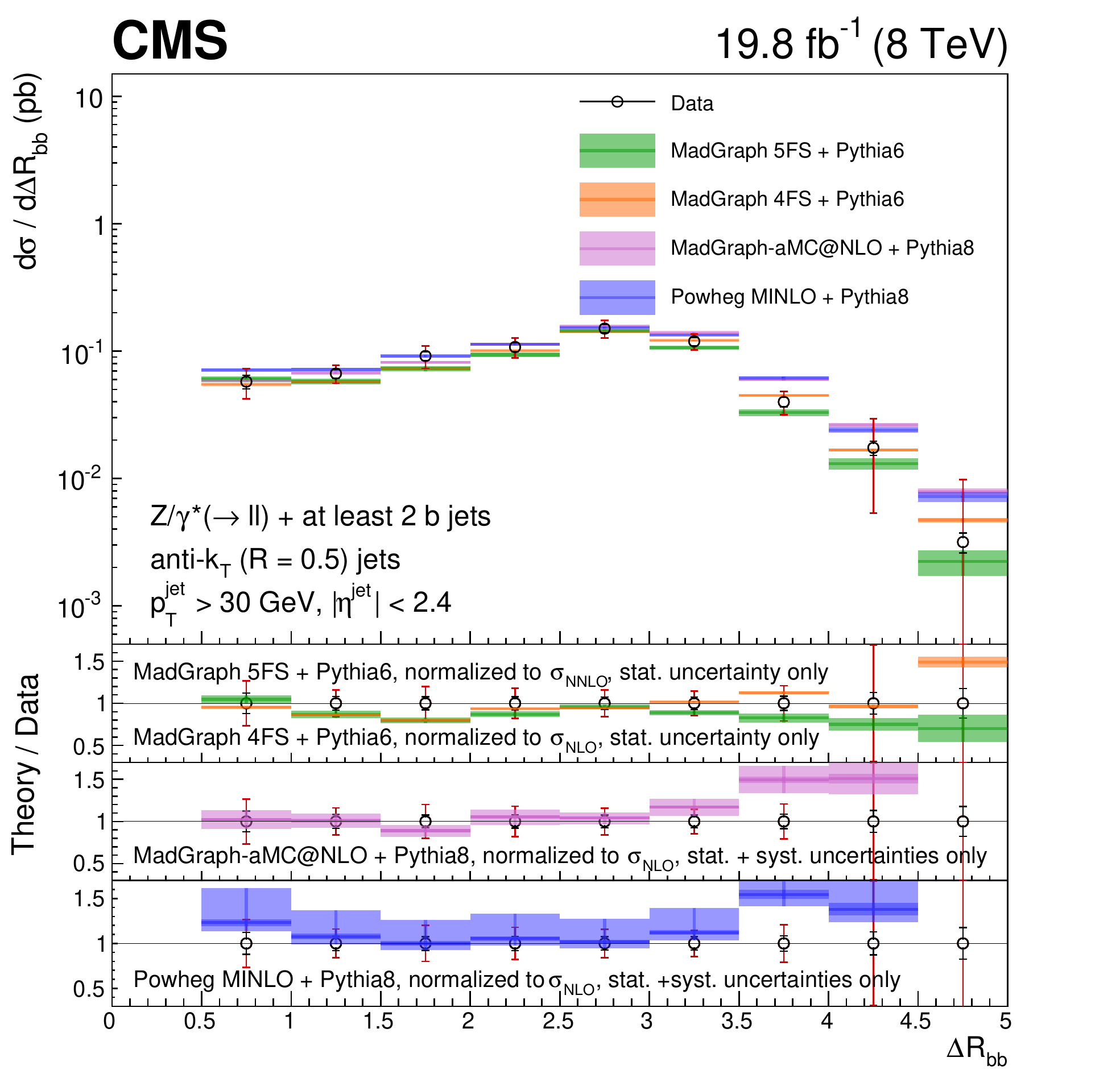}
\caption{Differential fiducial cross section for Z(2b) production as a
  function of $\Delta R_{\PQb\PQb}$, compared with the \MADGRAPH 5FS,
  \MADGRAPH 4FS, \AMCATNLO, and \POWHEG\ \textsc{minlo} theoretical
  predictions (shaded bands), normalized to the theoretical cross
  sections described in the text.  For each data point the statistical
  and the total (sum in quadrature of statistical and systematic)
  uncertainties are represented by the double error bar.  The width of
  the shaded bands represents the uncertainty in the theoretical
  predictions, and, for NLO calculations, theoretical systematic
  uncertainties are added in the ratio plots with the inner darker
  area representing the statistical component only.
\label{fig:w_DR_bb_unfolding}}
\end{figure}

\begin{figure}[hbt]
\centering
\includegraphics[width=\ghmFigWidthTwo]{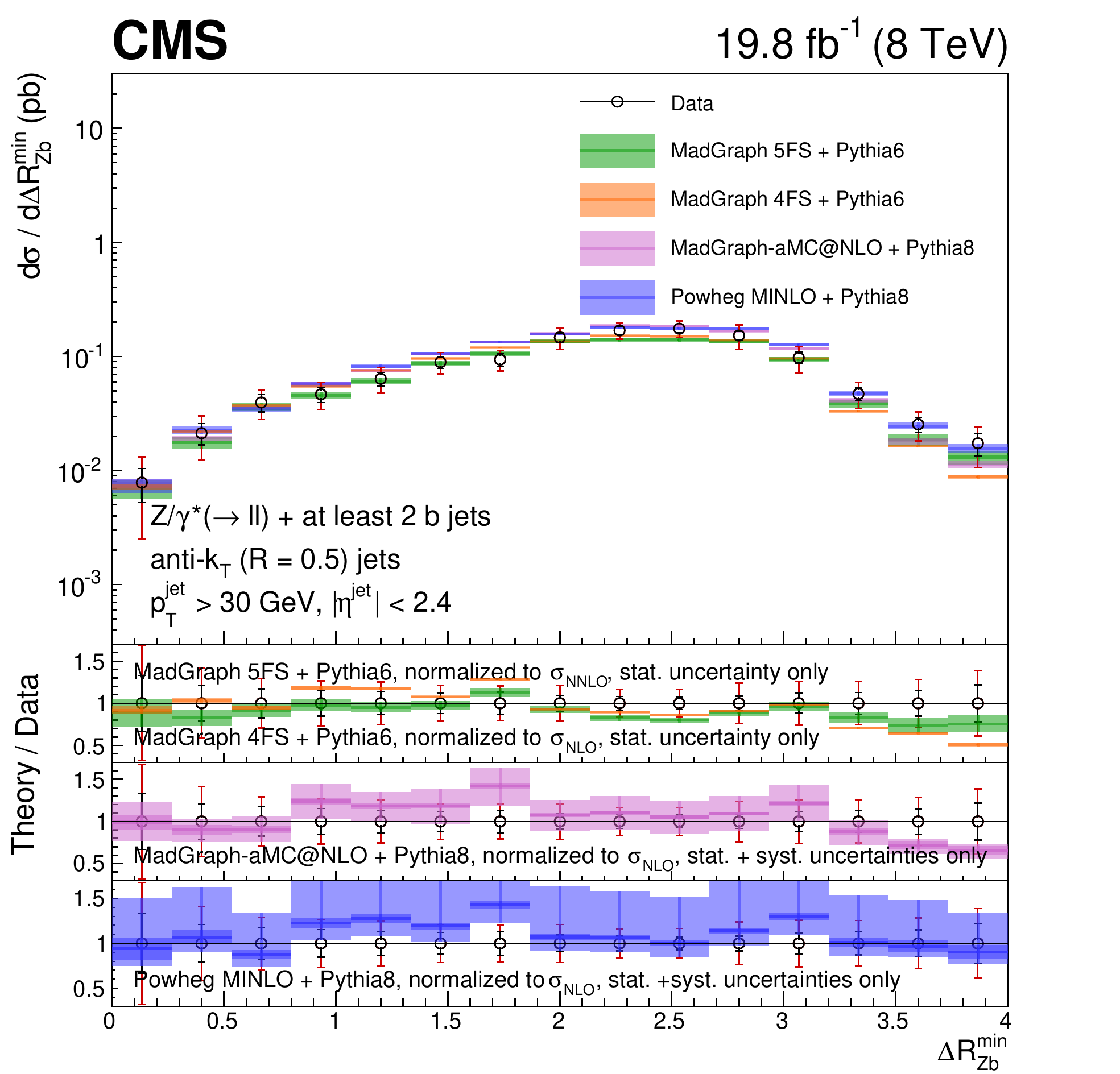}
\caption{Differential fiducial cross section for Z(2b) production as a
  function of $\Delta R_\PZ\PQb^\text{min}$, compared with the
  \MADGRAPH 5FS, \MADGRAPH 4FS, \AMCATNLO, and \POWHEG\ \textsc{minlo}
  theoretical predictions (shaded bands), normalized to the
  theoretical cross sections described in the text.  For each data
  point the statistical and the total (sum in quadrature of
  statistical and systematic) uncertainties are represented by the
  double error bar.  The width of the shaded bands represents the
  uncertainty in the theoretical predictions, and, for NLO
  calculations, theoretical systematic uncertainties are added in the
  ratio plots with the inner darker area representing the statistical
  component only.
\label{fig:w_DR_Zb_min_unfolding}}
\end{figure}

\begin{figure}[hbt]
\centering
\includegraphics[width=\ghmFigWidthTwo]{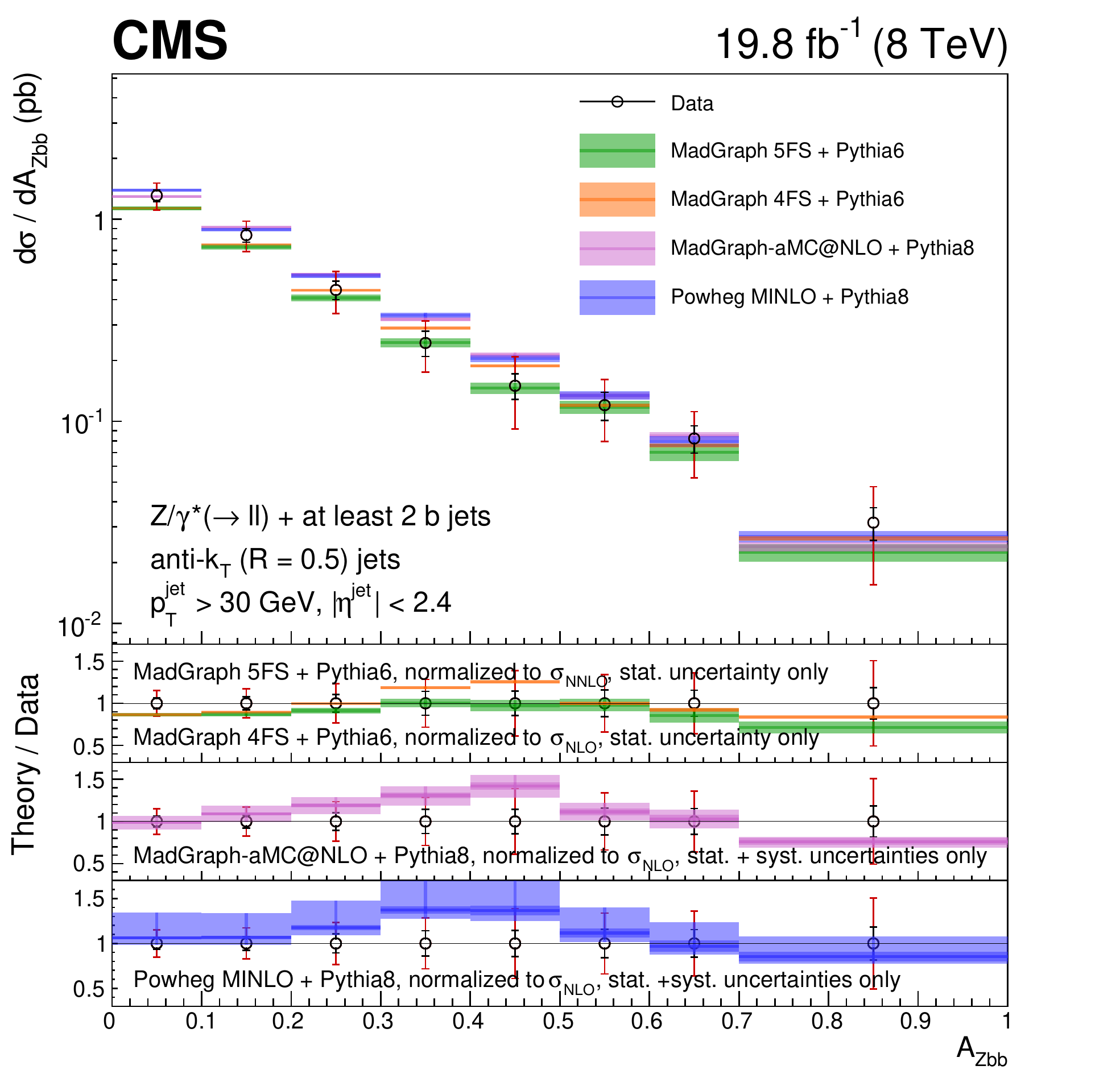}
\caption{Differential fiducial cross section for Z(2b) production as a
  function of $A_\PZ\PQb\PQb$, compared with the \MADGRAPH 5FS,
  \MADGRAPH 4FS, \AMCATNLO, and \POWHEG\ \textsc{minlo} theoretical
  predictions (shaded bands), normalized to the theoretical cross
  sections described in the text.  For each data point the statistical
  and the total (sum in quadrature of statistical and systematic)
  uncertainties are represented by the double error bar.  The width of
  the shaded bands represents the uncertainty in the theoretical
  predictions, and, for NLO calculations, theoretical systematic
  uncertainties are added in the ratio plots with the inner darker
  area representing the statistical component only.
\label{fig:w_A_Zb_unfolding}}
\end{figure}

The results for the differential cross sections measured with the \Ztb
event selection are shown in
Figs.~\ref{fig:w_first_bjet_pt_2b_unfolding}--\ref{fig:w_A_Zb_unfolding}.
Within uncertainties, no global normalization discrepancy is observed,
contrary to the \Zob case.  The leading and subleading jet spectra are
slightly underestimated in the soft region by the LO calculations (the
leading b jet \pt in the first two bins of
Fig.~\ref{fig:w_first_bjet_pt_2b_unfolding} and the subleading b jet
\pt in the first bin of Fig.~\ref{fig:w_second_bjet_pt_2b_unfolding}),
while the \PZ boson \pt distribution is well reproduced, within
uncertainties (Fig.~\ref{fig:w_pt_Z_b_2b_unfolding}). The 4FS
predictions overestimate the data at the high end of these \pt
distributions. The ratios of all theoretical predictions and the data
show a slight positive slope for the azimuthal separation
(Fig.~\ref{fig:w_delta_phi_2b_unfolding}). All the other distributions
are well reproduced. In general, given the experimental uncertainties,
the measurements do not strongly discriminate between the theoretical
predictions. The ratio of the \AMCATNLO and \POWHEG predictions based
on NLO matrix elements with data shows a similar shape.

\section{Summary}
The process of associated production of jets, including $\PQb$ jets,
and a \PZ boson decaying into lepton pairs ($\ell=\Pe,\mu$) are
measured in LHC $\Pp\Pp$ collisions at $\sqrt{s} = 8\TeV$ with the CMS
experiment, using a data set corresponding to an integrated luminosity
of 19.8\fbinv.  The measured fiducial cross sections are compared to
several theoretical predictions.  The cross sections are measured as a
function of various kinematic observables describing the event
topology with a \PZ boson and at least one $\PQb$ jet: the \pt and
$\eta$ of the leading $\PQb$ jet, the \PZ boson \pt, the scalar sum
\HT of the jet transverse momenta, and the azimuthal angular
difference between the directions of the leading $\PQb$ jet and the
\PZ boson.  A comparison is made of the unfolded data with
leading-order pQCD predictions based on matrix element calculations
matched with parton showering, testing models using the \MADGRAPH
event generator, or with the NLO calculations, merging predictions for
zero, one, and two extra jets with \AMCATNLO, or for the first two
jets with \POWHEG in the \textsc{minlo} approach. In most cases the
theoretical predictions agree with the data, although the
normalization for \MADGRAPH 4FS, which fails to describe simultaneously
both the low- and high-\pt $\cPqb$ jet regions, is underestimated by 20\%.
The ratios of differential cross sections for the production
of a \PZ boson in association with at least one $\PQb$ jet and the
inclusive $\PZ$+jets production are measured and compared with
theoretical expectations. The 4FS-based prediction fails to describe
the shape of the ratio as a function of the leading $\PQb$ jet \pt,
and discrepancies in the shape are also observed for high values of
the \PZ boson \pt.

The production of a \PZ boson in association with two $\PQb$ jets is
also investigated.  In this case the kinematic observables are the
transverse momenta of the leading and subleading $\PQb$ jets, the \pt
of the \PZ boson, the separations of the $\PQb$ jets both in azimuthal
angle and in the $\eta$-$\phi$ plane, the minimal distance in the
$\eta$-$\phi$ plane between the \PZ boson and a $\PQb$ jet, the
asymmetry between the minimal and the maximal distances between the \PZ
boson and a $\PQb$ jet, and the invariant masses of the $\PQb\PQb$
and the $\PZ\PQb\PQb$ systems.  The measured distributions are
generally well reproduced by the predictions.

\clearpage
\begin{acknowledgments}

\hyphenation{Bundes-ministerium Forschungs-gemeinschaft Forschungs-zentren Rachada-pisek} We congratulate our colleagues in the CERN accelerator departments for the excellent performance of the LHC and thank the technical and administrative staffs at CERN and at other CMS institutes for their contributions to the success of the CMS effort. In addition, we gratefully acknowledge the computing centres and personnel of the Worldwide LHC Computing Grid for delivering so effectively the computing infrastructure essential to our analyses. Finally, we acknowledge the enduring support for the construction and operation of the LHC and the CMS detector provided by the following funding agencies: the Austrian Federal Ministry of Science, Research and Economy and the Austrian Science Fund; the Belgian Fonds de la Recherche Scientifique, and Fonds voor Wetenschappelijk Onderzoek; the Brazilian Funding Agencies (CNPq, CAPES, FAPERJ, and FAPESP); the Bulgarian Ministry of Education and Science; CERN; the Chinese Academy of Sciences, Ministry of Science and Technology, and National Natural Science Foundation of China; the Colombian Funding Agency (COLCIENCIAS); the Croatian Ministry of Science, Education and Sport, and the Croatian Science Foundation; the Research Promotion Foundation, Cyprus; the Secretariat for Higher Education, Science, Technology and Innovation, Ecuador; the Ministry of Education and Research, Estonian Research Council via IUT23-4 and IUT23-6 and European Regional Development Fund, Estonia; the Academy of Finland, Finnish Ministry of Education and Culture, and Helsinki Institute of Physics; the Institut National de Physique Nucl\'eaire et de Physique des Particules~/~CNRS, and Commissariat \`a l'\'Energie Atomique et aux \'Energies Alternatives~/~CEA, France; the Bundesministerium f\"ur Bildung und Forschung, Deutsche Forschungsgemeinschaft, and Helmholtz-Gemeinschaft Deutscher Forschungszentren, Germany; the General Secretariat for Research and Technology, Greece; the National Scientific Research Foundation, and National Innovation Office, Hungary; the Department of Atomic Energy and the Department of Science and Technology, India; the Institute for Studies in Theoretical Physics and Mathematics, Iran; the Science Foundation, Ireland; the Istituto Nazionale di Fisica Nucleare, Italy; the Ministry of Science, ICT and Future Planning, and National Research Foundation (NRF), Republic of Korea; the Lithuanian Academy of Sciences; the Ministry of Education, and University of Malaya (Malaysia); the Mexican Funding Agencies (BUAP, CINVESTAV, CONACYT, LNS, SEP, and UASLP-FAI); the Ministry of Business, Innovation and Employment, New Zealand; the Pakistan Atomic Energy Commission; the Ministry of Science and Higher Education and the National Science Centre, Poland; the Funda\c{c}\~ao para a Ci\^encia e a Tecnologia, Portugal; JINR, Dubna; the Ministry of Education and Science of the Russian Federation, the Federal Agency of Atomic Energy of the Russian Federation, Russian Academy of Sciences, and the Russian Foundation for Basic Research; the Ministry of Education, Science and Technological Development of Serbia; the Secretar\'{\i}a de Estado de Investigaci\'on, Desarrollo e Innovaci\'on and Programa Consolider-Ingenio 2010, Spain; the Swiss Funding Agencies (ETH Board, ETH Zurich, PSI, SNF, UniZH, Canton Zurich, and SER); the Ministry of Science and Technology, Taipei; the Thailand Center of Excellence in Physics, the Institute for the Promotion of Teaching Science and Technology of Thailand, Special Task Force for Activating Research and the National Science and Technology Development Agency of Thailand; the Scientific and Technical Research Council of Turkey, and Turkish Atomic Energy Authority; the National Academy of Sciences of Ukraine, and State Fund for Fundamental Researches, Ukraine; the Science and Technology Facilities Council, UK; the US Department of Energy, and the US National Science Foundation.

Individuals have received support from the Marie-Curie programme and the European Research Council and EPLANET (European Union); the Leventis Foundation; the A. P. Sloan Foundation; the Alexander von Humboldt Foundation; the Belgian Federal Science Policy Office; the Fonds pour la Formation \`a la Recherche dans l'Industrie et dans l'Agriculture (FRIA-Belgium); the Agentschap voor Innovatie door Wetenschap en Technologie (IWT-Belgium); the Ministry of Education, Youth and Sports (MEYS) of the Czech Republic; the Council of Science and Industrial Research, India; the HOMING PLUS programme of the Foundation for Polish Science, cofinanced from European Union, Regional Development Fund, the Mobility Plus programme of the Ministry of Science and Higher Education, the National Science Center (Poland), contracts Harmonia 2014/14/M/ST2/00428, Opus 2013/11/B/ST2/04202, 2014/13/B/ST2/02543 and 2014/15/B/ST2/03998, Sonata-bis 2012/07/E/ST2/01406; the Thalis and Aristeia programmes cofinanced by EU-ESF and the Greek NSRF; the National Priorities Research Program by Qatar National Research Fund; the Programa Clar\'in-COFUND del Principado de Asturias; the Rachadapisek Sompot Fund for Postdoctoral Fellowship, Chulalongkorn University and the Chulalongkorn Academic into Its 2nd Century Project Advancement Project (Thailand); and the Welch Foundation, contract C-1845.

\end{acknowledgments}
\bibliography{auto_generated}
\cleardoublepage \appendix\section{The CMS Collaboration \label{app:collab}}\begin{sloppypar}\hyphenpenalty=5000\widowpenalty=500\clubpenalty=5000\textbf{Yerevan Physics Institute,  Yerevan,  Armenia}\\*[0pt]
V.~Khachatryan, A.M.~Sirunyan, A.~Tumasyan
\vskip\cmsinstskip
\textbf{Institut f\"{u}r Hochenergiephysik,  Wien,  Austria}\\*[0pt]
W.~Adam, E.~Asilar, T.~Bergauer, J.~Brandstetter, E.~Brondolin, M.~Dragicevic, J.~Er\"{o}, M.~Flechl, M.~Friedl, R.~Fr\"{u}hwirth\cmsAuthorMark{1}, V.M.~Ghete, C.~Hartl, N.~H\"{o}rmann, J.~Hrubec, M.~Jeitler\cmsAuthorMark{1}, A.~K\"{o}nig, M.~Krammer\cmsAuthorMark{1}, I.~Kr\"{a}tschmer, D.~Liko, T.~Matsushita, I.~Mikulec, D.~Rabady, N.~Rad, B.~Rahbaran, H.~Rohringer, J.~Schieck\cmsAuthorMark{1}, J.~Strauss, W.~Treberer-Treberspurg, W.~Waltenberger, C.-E.~Wulz\cmsAuthorMark{1}
% ...
\vskip\cmsinstskip
% ...
1:~~Also at Vienna University of Technology, Vienna, Austria\\
% ...

\end{sloppypar}
\end{document}